\documentclass[sigplan,twocolumn]{acmart}
\renewcommand\footnotetextcopyrightpermission[1]{}
\usepackage[]{hyperref}

\usepackage[normalem]{ulem}
\usepackage{textgreek}
\usepackage{textcomp}

\usepackage{multirow}
\usepackage{xspace}
\usepackage{subfig}
\usepackage[font=small,labelfont=bf]{caption}
\usepackage{graphicx}

\usepackage{fancyhdr}
\usepackage{colortbl}
\usepackage{tabulary}
\usepackage{etoolbox}
\usepackage{booktabs}
\usepackage{mathtools}
\usepackage{pifont}
\usepackage{color}

\newcommand{\fig}[1]{Figure~\ref{#1}}
\newcommand{\sect}[1]{Section~\ref{#1}}
\newcommand{\tab}[1]{Table~\ref{#1}}

\newcommand{\miglarge}[0]{7g.40gb(1x)\xspace}
\newcommand{\migsmall}[0]{1g.5gb(7x)\xspace}
\newcommand{\migmed}[0]{2g.10gb(3x)\xspace}
\newcommand{\proposed}[0]{PREBA\xspace}

\newcommand{\batchMax}[0]{$Batch_{max}$\xspace}
\newcommand{\batchQueue}[0]{$Time_{queue}$\xspace}
\newcommand{\batchKnee}[0]{$Batch_{knee}$\xspace}
\newcommand{\tailKnee}[0]{$Time_{knee}$\xspace}

\begin{document}

\title{\proposed: A Hardware/Software Co-Design for\\
Multi-Instance GPU based AI Inference Servers
}

\author{Gwangoo Yeo}
\authornote{Co-first authors who contributed equally to this research.}
\affiliation{%
    \institution{KAIST}
    \country{}
}
\email{gwangoo525@kaist.ac.kr}
\author{Jiin Kim}
\authornotemark[1]
\affiliation{%
    \institution{KAIST}
    \country{}
}
\email{jiin.kim@kaist.ac.kr}
\author{Yujeong Choi}
\affiliation{%
    \institution{Google}
    \country{}
}
\email{yujeong.choi.0606@gmail.com}
\author{Minsoo Rhu}
\affiliation{%
    \institution{KAIST}
    \country{}
}
\email{mrhu@kaist.ac.kr}
\setlength{\textfloatsep}{8pt}

\begin{abstract}

NVIDIA's Multi-Instance GPU (MIG) is a
feature that enables system designers to reconfigure one large GPU
into multiple smaller GPU slices. This work 
characterizes this emerging GPU and evaluates its effectiveness in designing
high-performance AI inference servers. Our study reveals that the data
preprocessing stage of AI inference causes significant performance bottlenecks
to MIG. To this end, we
present \proposed, which is a hardware/software co-design targeting MIG
inference servers. Our first proposition is an FPGA-based data preprocessing
accelerator that unlocks the full potential of MIG with domain-specific
acceleration of data preprocessing. The MIG inference server unleashed from
preprocessing overheads is then augmented with our dynamic batching system that
enables high-performance inference. PREBA is implemented
end-to-end in real systems, providing 
$3.7\times$ improvement in throughput, $3.4\times$ reduction
in tail latency, $3.5\times$ improvement in energy-efficiency, and $3.0\times$ improvement
in cost-efficiency.

\end{abstract}

\maketitle 

\pagestyle{plain} 

\section{Introduction}
\label{sect:intro}

Unlike throughput-hungry AI training algorithms, which is well-suited for acceleration using throughput-optimized
GPUs, fully saturating a GPU's high compute power and memory bandwidth under small batch inference
scenarios is much more challenging.
From an Artificial Intelligence as a Service (AIaaS) provider's perspective, maintaining high
GPU resource utilization is of utmost importance to optimize Total Cost of Ownership (TCO). As such, 
the low resource utilization in GPU-based AI inference presents an important research challenge
with significant industrial importance.

To overcome such challenge, recent high-end GPUs from NVIDIA are equipped with
a feature called \emph{Multi-Instance GPU} (MIG)~\cite{mig} which allows one
large GPU's resources to be partitioned into multiple
smaller sized GPU \emph{slices}. Each GPU slice (henceforth referred to as a
		\emph{virtual GPU}, vGPU) functions as a standalone GPU and can
independently be handed over to a Virtual Machine (VM) with performance
isolation guarantees using Single Root I/O Virtualization (SR-IOV~\cite{sr_iov}). 
		Such reconfigurability becomes extremely
valuable for AIaaS providers because GPUs can now be deployed for both
training (configured as one large monolithic GPU) and inference (partitioned
		into multiple vGPUs) with high GPU utilization.  For instance, each vGPU
can independently host an inference server and exploit query-level parallelism
to improve GPU's utilization, e.g., a single NVIDIA A100 GPU~\cite{a100} can be
partitioned into seven small vGPUs and host seven inference servers to
concurrently handle multiple service queries.

Given this landscape, a key objective of this paper is to characterize this
emerging, reconfigurable GPU architecture and evaluate the efficacy of
MIG for latency-critical AI workloads. 
Although we confirm that
an MIG inference server partitioned into multiple small vGPUs is indeed highly
effective in improving GPU utilization, we observe that the  \emph{data
preprocessing} stage of inference incurs a critical performance bottleneck
when a GPU is partitioned using MIG.  Current AI inference servers utilize
the CPU to apply various application-specific preprocessing to the raw input
data before sending them to the GPU for AI model execution (e.g., image
decoding, resizing, cropping, and others for computer vision~\cite{pytorch_imagepreproc}).
Our key observation is that the overhead of CPU-side data preprocessing
increases proportionally to the number of vGPUs instantiated, causing
significant reduction in Queries Processed per Second (QPS) vs. an inference
server without any data preprocessing overheads. 

Another important research challenge with MIG is how to best exploit the
multitude of vGPUs available in the inference server.  Concretely, compared
to the baseline system where one large GPU is available for scheduling
inputs, a MIG inference server must consider multiple smaller vGPUs for
scheduling.  A critical aspect of input scheduling is how it should
\emph{batch} multiple inputs to improve GPU utilization. 
We observe that the batching algorithm, when naively implemented for MIG inference servers, can suffer from high tail latency even at small batch sizes.
Therefore, the batching system must consider MIG's unique properties for optimizing input batching and scheduling decisions, one which to the best of our knowledge prior work has neglected upon.

To this end, we propose \proposed ({\bf PRE}processing and {\bf BA}t-\\ching system) which is a hardware/software co-design for high-performance MIG inference servers.
The key contributions of our proposed system are twofold:

\begin{enumerate}
\item {\bf (Hardware) A Data Processing Unit (DPU) for MIG.} \proposed
fundamentally addresses MIG's CPU-side data preprocessing bottlenecks by
completely offloading performance-critical data preprocessing operations to an
FPGA-based DPU.  Because of the diversity of AI inference workloads, it is
important that \proposed's hardware architecture contains enough flexibility to
handle various application-specific data preprocessing operations. We show that
our FPGA-based DPU can seamlessly accelerate the data preprocessing operations
of diverse AI workloads.  Our DPU microarchitecture is co-designed with the
inference server's input batching system so that it is tuned for the
latency-critical nature of AI inference. Specifically, our DPU is optimized for
minimizing a \emph{single}-input request's latency while maximizing throughput,
 enabling our batching system to flexibly adjust the batching granularity based on the vGPU size, the AI model, and input size/length at runtime.

\item {\bf (Software) A dynamic batching system for MIG.} Designing an
efficient batching system requires careful tuning  of the following two
hyperparameters: (a) the \emph{maximum batch size} (\batchMax, i.e., the
batching system will schedule the batched inputs once a certain number of
inputs are ready for batched execution) and (b) the \emph{maximum queueing delay for
batching} (\batchQueue, i.e., the longest time period the batching system
will have input requests wait inside a queue to form a larger batch). We
observe that tuning the values of these two hyperparameters for a given AI
model without considering MIG's effect on batching and scheduling (e.g., the sensitivity of a given vGPU's model execution time and throughput as a function of batch size) leads to sub-optimal performance. 			
We propose a dynamic batching system that leverages a profiling-based analytical model to systematically estimate the optimal \batchMax and \batchQueue values to utilize for high-performance MIG inference servers.

\end{enumerate}

Overall, \proposed presents a practical yet highly effective architectural
solution targeting MIG.  To the best of our knowledge, this
work is the first to identify, analyze, and explore the data preprocessing
bottlenecks and the need for an efficient batching system tailored for
inference servers utilizing MIG, an emerging yet highly important architectural feature
introduced in state-of-the-art high-end NVIDIA GPUs.  
\proposed is implemented end-to-end over real systems using commodity hardware and open-source software, providing an average $3.7$ $\times$ improvement in throughput, $3.4\times$ reduction in tail latency, $3.5\times$ improvement in energy-efficiency, and $3.0\times$ improvement in cost-efficiency.

\section{Background}
\label{sect:background}
\label{page:time_queue}

\subsection{AI Training vs. Inference in GPUs}
\label{sect:background_inf_in_gpus}

GPUs are throughput-optimized processors employing many-core SIMD vectors
with high-bandwidth memory (e.g., HBM\\~\cite{hbm3e} or GDDR~\cite{gddr7}).  
Because AI training tasks employ large input batch sizes
and exhibit throughput-hungry characteristics, GPUs have become the de facto
standard in designing today's AI training systems.
For AI inference, however, the input batch size is orders of magnitude
smaller than training. Therefore, the working set of inference is typically
too small to fully saturate a high-end GPU's compute and memory resources. 
As such,
GPU-based inference servers oftentimes experience very low GPU utilization. To this end, there has been significant interest from both 
academia~\cite{cnvlutin,song:2015:eie,eyeriss,scnn,ankit2019puma,bitfusion:isca:2018,laconic:isca:2019, Neuralcache:isca:2018, Simba:micro:2019,MAICC:micro:2023,nonesparse:dac:2019,realtime:dac:2017}
and industry~\cite{tpu_paper,inferentia,Maia,mita,rebellions,graphcore} to design a domain-specific architecture targeting AI inference. 
Nonetheless, outside of some hyperscalers~\cite{tpu_paper,inferentia,Maia,mita}, the AI inference market is still dominated by GPUs as they are readily available with their mature software stack, functioning as the ``go-to'' platform for deploying AI services.

Despite GPU's prominence in deploying AI services, the GPU underutilization issue upon small batch inference scenarios is still a weak spot of GPU-based AI inference servers.
To address such limitation, GPU vendors have released  small, inference-purposed
GPUs that contain relatively lower compute and memory throughput (e.g., NVIDIA T4~\cite{t4}), allowing better GPU utility even for small batch inference.
Unfortunately, deploying these small GPUs for inference introduces a significant tradeoff: it reduces the \emph{computational
density} of the inference server, proportional to the performance difference between
a large (training-purposed) and small (inference-purposed) GPU. NVIDIA's recently announced
\emph{Multi-Instance GPU} (MIG~\cite{mig}) architecture is intended to remedy the aforementioned
challenges, allowing a GPU card to be (re)configured as either a single large GPU or be partitioned into
multiple small GPU slices (referred to as \emph{virtual GPUs}, vGPUs, in the rest of this paper).  In the following section, we delve deeper into the
MIG architecture.

\subsection{NVIDIA's MIG with Reconfigurability}
\label{sect:background_mig}

\begin{figure}[t!] \centering
  \includegraphics[width=0.47\textwidth]{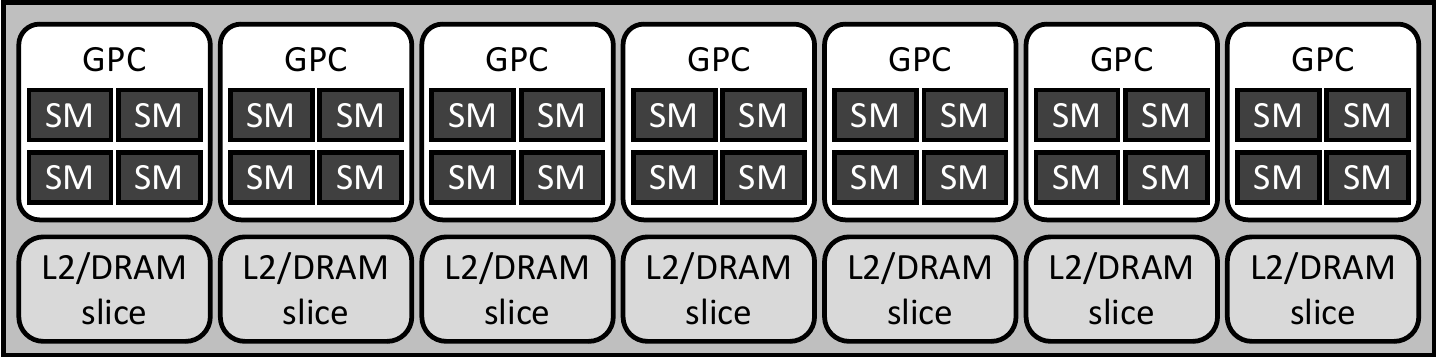} 
\vspace{-0.4em}
  \caption{Overview of NVIDIA's GPU architecture.}
  \label{fig:gpu_arch}
\end{figure}

\begin{figure}[t!] \centering
  \includegraphics[width=0.47\textwidth]{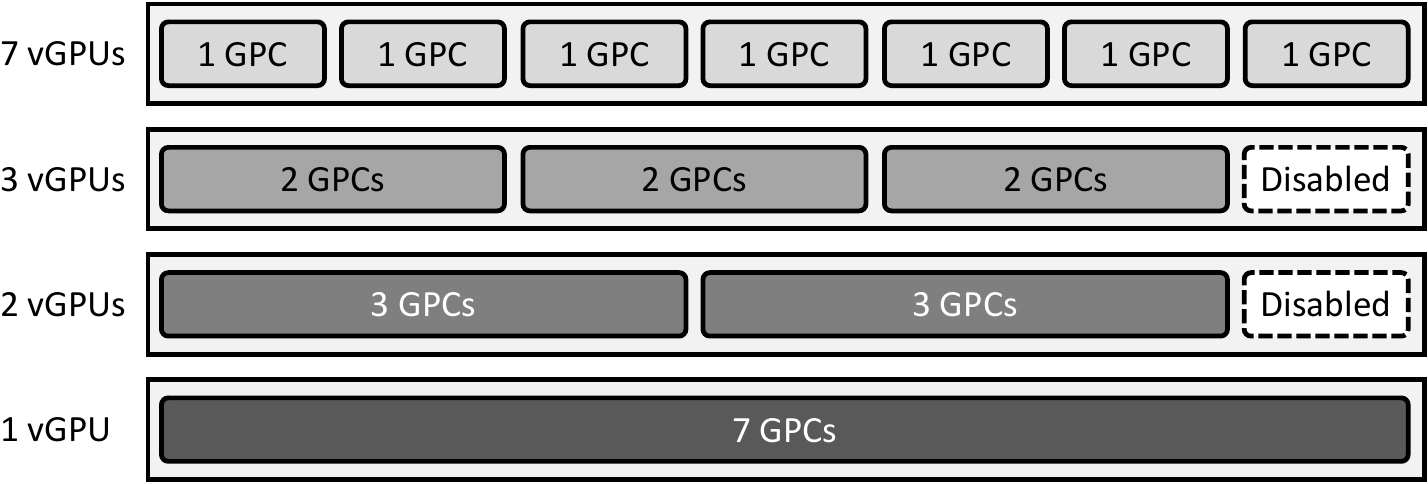}
\vspace{-0.4em}
  \caption{MIG partitioning options in NVIDIA A100 GPU.}
  \label{fig:gpu_mig}
\end{figure}

\begin{sloppypar}\tolerance 900
This work employs NVIDIA's MIG as the foundation for building an AI inference
server. Below we elaborate on NVIDIA's A100
GPU~\cite{a100} which features MIG capabilities. 
\end{sloppypar}

\textbf{GPU hardware architecture.} 
\fig{fig:gpu_arch} provides a high-level overview of a modern GPU architecture from NVIDIA.
A Streaming Multiprocessor (SM) is the most fundamental building block in an NVIDIA GPU,
which is essentially a SIMD  processor~\cite{cuda}.  Each SM
includes a large register-file, a scratchpad, and an L1 cache
to capture locality.  A group of SMs constitute a cluster, which is
called a Graphics Processing Cluster (GPC), and the SMs within a GPC
share the communication channel to the Network-on-Chip (NoC).  The NoC
is implemented using a crossbar which enables GPCs to access
the L2 cache and the corresponding DRAM
channel. 

\textbf{Multi-Instance GPU.} NVIDIA's MIG is architected using the GPCs
(computation) and L2/DRAM slices (memory) as the
basic building block. A vGPU (a GPU slice) is defined in the granularity of
a GPC and an A100 GPU containing seven GPCs can be configured up to seven vGPUs,
each containing a single GPC. There are restrictions on how many L2/DRAM slices
a vGPU can be allocated with (e.g., it is impossible to combine a single GPC with
four L2/DRAM slices into a single vGPU), so a vGPU must be configured within
the limited set of possible ``GPC $\times$ L2/DRAM'' combinations dictated by NVIDIA. 
\fig{fig:gpu_mig} shows possible MIG configurations in NVIDIA A100 and how they
can be utilized to instantiate multiple vGPUs, each running an AI inference server.

\begin{figure}[t!] \centering
  \includegraphics[width=0.47\textwidth]{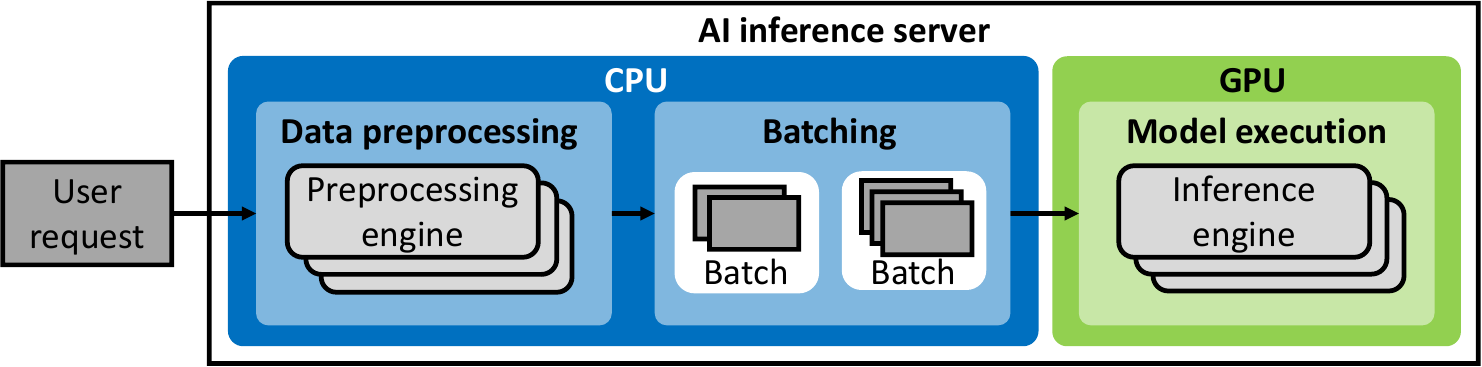} 
\vspace{-0.4em}
  \caption{End-to-end AI inference pipeline. }
  \label{fig:inference_pipeline}
\end{figure}

\subsection{AI Model Serving Pipeline for Inference}
\label{sect:background_serving}

\fig{fig:inference_pipeline} illustrates the end-to-end AI model serving
pipeline to service an inference query (e.g., NVIDIA Triton Inference
		Server~\cite{trtis}).  As depicted, the first two stages of model serving
(data preprocessing, batching) are executed on the CPU while the GPU executes
the last stage (model execution).  Below we detail the operations conducted in
each stage.

\begin{figure}[t!] \centering
\vspace{-1.1em}
\subfloat[]{\includegraphics[width=0.32\textwidth]{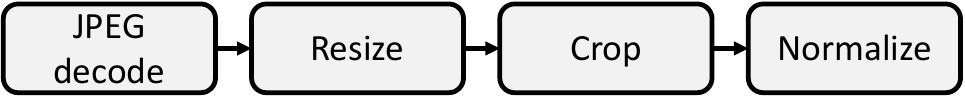}}\\
\vspace{-0.4em}
\subfloat[]{\includegraphics[width=0.47\textwidth]{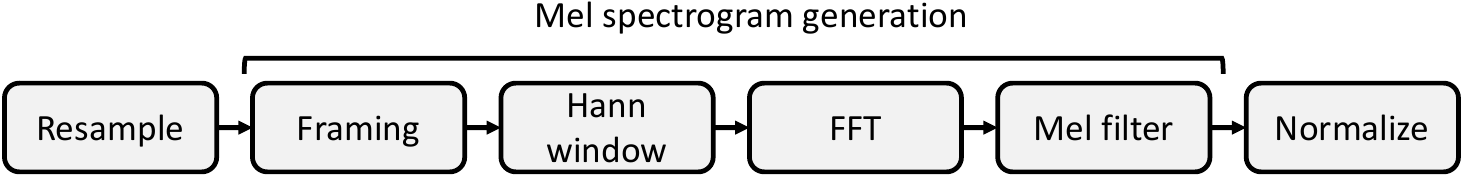}}
\vspace{-0.4em}
\caption{Data preprocessing operations for (a) computer vision and (b) audio processing.}
\label{fig:data_preprocessing}
\end{figure}

\begin{enumerate}
\item {\bf Data preprocessing.} The raw input data routed to the inference server is first
preprocessed in order to be transformed into a data format that is compatible with the AI model. For the computer vision
and audio processing AI workloads we focus on in this work, some representative preprocessing operations include a) decoding (JPEG$\rightarrow$RGB),
resizing, and cropping to generate the preprocessed image (e.g., 224$\times$224$\times$3) for computer vision~\cite{pytorch_imagepreproc}, and 
		b) resampling, Fast Fourier Transform (FFT), and applying Mel filters for audio processing~\cite{pytorch_audiopreproc} (\fig{fig:data_preprocessing}).

\item {\bf Batching.} Throughput-optimized
GPUs prefer large input tensors over smaller ones to maximally exploit its abundant compute and memory performance (\sect{sect:background_inf_in_gpus}).
Therefore, coalescing multiple preprocessed inputs into a single large input batch plays
a critical role in achieving high GPU utilization~\cite{cellular_batching,lazybatching,yu2022orca}. Another key benefit of constructing a large input batch is that
it increases the reuse of AI model parameters uploaded from the off-chip DRAM during the model execution stage, significantly improving GPU utility.

\item {\bf Model execution.} The final stage of model serving is the actual execution of the Deep Neural Network (DNN) layers
that define the AI model. Popular inference serving systems like NVIDIA Triton Inference Server~\cite{trtis} or TensorFlow Serving~\cite{tf_serving}
utilize an \emph{inference execution engine} to perform the tensor operations that constitute the DNN layer (e.g., TensorRT~\cite{tensorrt}).
Once the final model output is derived, the CPU retrieves the final result and returns it back to the client.
\end{enumerate}

\subsection{Related Work}
\label{sect:related}

{\bf Data preprocessing for AI training/inference.} Mohan et
al.~\cite{mohan2021analyzing} analyzed the impact of input data preprocessing
on end-to-end AI training's throughput, proposing various performance
optimization strategies like I/O caching and sharing preprocessed training
dataset across several training hyperparameter search jobs to amortize the data
preprocessing overhead. Trainbox~\cite{park2020trainbox} similarly observes data
preprocessing bottlenecks in AI training tasks and proposes to offload the
compute-intensive data preparation to an FPGA device. There is also a line of
work that proposes a disaggregated pool of CPU compute nodes, dedicated for data preprocessing, and allocates
them on-demand per the training task's preprocessing throughput
requirements~\cite{zhao2022dsi,graur2022cachew,um2023fastflow}. The work from Zhao et
al.~\cite{zhao2022dsi}, for instance, puts a particular emphasis on optimizing the
data preprocessing stage of recommendation model training.
Similar to Trainbox, DLBooster~\cite{cheng2019dlbooster} seeks to address the data
preprocessing bottlenecks of AI by offloading
preprocessing operations to an FPGA device.  None of these prior art explores
the implication of data preprocessing overheads over NVIDIA's MIG
nor proposes latency-optimization strategies tailored for data preprocesing
accelerators.
 Overall, the key contribution of our work is orthogonal to these related work
as our study is focused on AI inference servers employing NVIDIA's reconfigurable MIG
architecture.

{\bf Exploration of NVIDIA MIG.} PARIS and ELSA~\cite{paris_and_elsa} 
explored NVIDIA's MIG architecture for AI
inference, proposing optimal MIG partitioning strategies as well as
high-performance scheduling policies for inference. However, this work assumes
that the data preprocessing stage is always able to supply sufficient amount of
preprocessed inputs to the model execution stage. Consequently, Kim et
al.~\cite{paris_and_elsa} only focuses on the backend GPU-side inference while 
data preprocessing is disabled (\fig{fig:inference_pipeline}).  MISO~\cite{miso}
explores the best MIG partitioning strategies for co-located AI training jobs in a
multi-tenant GPU cluster environment. Li et al.~\cite{li2022characterizing} and Robroek et
al.~\cite{robroek2022analysis} are also recent studies that characterize the performance of MIG
for AI workloads.  None of these prior work analyzes the data preprocessing
bottlenecks in a MIG inference server let alone efficient batching strategies for MIG, rendering the key contribution of our
work unique.

 {\bf Software solutions to improve GPU utilization in latency-critical applications.} 
LAX~\cite{LAX}, BayMax~\cite{Baymax:2016}, and Prophet~\cite{Prophet:2017}
develop solutions to improve GPU's resource utilization
without compromising Quality of Service (QoS) for latency-critical workloads. 
For instance, LAX points out the latency overheads of batching for latency-sensitive GPU applications and proposes to run multiple GPU tasks concurrently within a \emph{heterogeneous} set of GPU tasks (including AI workloads) 
using GPU streams, instead of batching.
Baymax and Prophet seek to improve GPU utilization by estimating resource contention at shared resources (e.g., PCIe bandwidth) and utilize
that information to intelligently determine which GPU tasks to concurrently execute among a \emph{heterogeneous} set of GPU workloads (a combination of generic GPU workloads like Rodinia~\cite{rodinia} and AI workloads~\cite{djinn_and_tonic}). While not specifically targeting GPUs,
		 PREMA~\cite{prema} seeks to improve an AI accelerator's resource utility using a preemptive scheduling algorithm targeting
		 a heterogeneous set of AI workloads (but without concurrent task execution).
A key distinction from these prior work vs. PREBA's batching system is that our solution employs batching input requests targeting the \emph{same} AI model and executes the batched input \emph{synchronously} to improve GPU's throughput, unlike
	these prior work seeking to \emph{asynchronously} execute multiple GPU workloads simultaneously among a heterogeneous set of GPU tasks. Another key distinction of PREBA is that our dynamic batching system is co-designed with our frontend DPU architecture tailored to address the unique challenges of NVIDIA's MIG system.
 BatchMaker~\cite{Batchmaker} 
and LazyBatching~\cite{lazybatching} suggest a fine-grained batching algorithm that coalesces inputs at the granularity of individual graph
nodes (e.g., cells in RNN models or layers in DNN models) to improve batching efficiency and thus enhance GPU utilization. A key focus of PREBA is the identification of the unique challenges MIG's batching system faces (i.e., its much lower \batchKnee, the maximum batch size at the knee of the tail latency curve) and how to address them by co-designing it with the frontend data preprocessing stage. As the key insights behind BatchMaker and LazyBatching are orthogonal to PREBA's batching system, they can be employed on top of our proposed system to further enhance performance.

{\bf Accelerating system software using DPUs.} While not targeting AI, there exists prior literature from academia~(e.g.,
TCP/IP~\cite{boo:2023:f4t, moon2020acceltcp, arashloo2020enabling,
shashidhara2022flextoe, tork2020lynx}, I/O virtualization~\cite{kwon:2020:osdi,
kwon:2021:atc, li2020leapio, chen2023bm}, P2P data movements~\cite{dcs_ctrl,
dcs, wang2022fpganic}, remote I/O access over the network~\cite{kim2023rearchitecting,
min2021gimbal}) and commercial products from industry (e.g., NVIDIA's
BlueField~\cite{bluefield}, Intel's IPU~\cite{intel_ipu}) that seek to offload CPU-intensive system software
 to an I/O attached accelerator~\cite{dcs,dcs_ctrl,boo:2023:f4t,kwon:2020:osdi,kwon:2021:atc,bluefield,intel_ipu,mangoboost,fungible,amazon_nitro}, similar in spirit to \proposed's DPU.  These latest research trends highlight the practicality and timeliness of our proposal.
\section{Characterization}
\label{sect:characterization}
\label{page:novelty_batch}

This section utilizes NVIDIA's A100 GPU to conduct a characterization of MIG inference servers. For brevity, we refer to a MIG partitioned into $V$ vGPUs where each vGPU contains $M$ GPCs (compute) and $N$ GB of DRAM (memory) as a ``Mg.Ngb(Vx)'' configuration\footnote{In this section, we evaluate \migsmall, \migmed, and \miglarge. Because \migmed only utilizes a total of 6 GPCs (NVIDIA prevents the remaining 1 GPC from being activated, see \fig{fig:gpu_mig}), the maximum throughput provided with \migmed is $14.2\%$ (=1/7) smaller than \migsmall and \miglarge. We nonetheless present our characterization study under this design point to highlight the implication of MIG's partitioning granularity on various aspects of AI inference servers.}. For instance, an A100 GPU
partitioned into seven vGPUs is \migsmall, whereas one without any partitioning and functions as one big vGPU is \miglarge (\fig{fig:gpu_mig}). As discussed later using \fig{fig:librispeech_dataset}, the size of a single-input for an audio processing application can vary, unlike the fixed-size input for computer vision workloads. For brevity, all experiments discussed in this section assume that the input audio length is fixed at $2.5$ sec. Nonetheless, we emphasize that the key observations discussed in this section remain intact regardless of the audio length. \sect{sect:methodology} further details our evaluation methodology.

\begin{figure}[t!] \centering
\subfloat[Computer vision]{\includegraphics[width=0.48\textwidth]{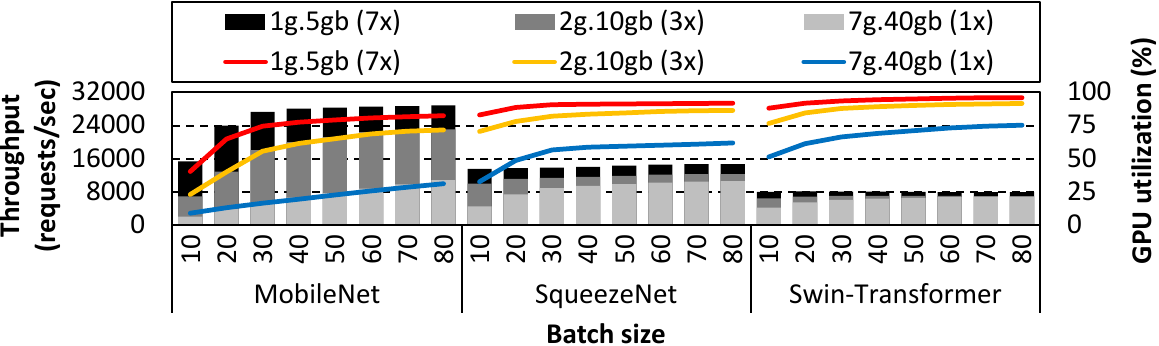}}\\
\vspace{-0.4em}
\subfloat[Audio processing]{\includegraphics[width=0.48\textwidth]{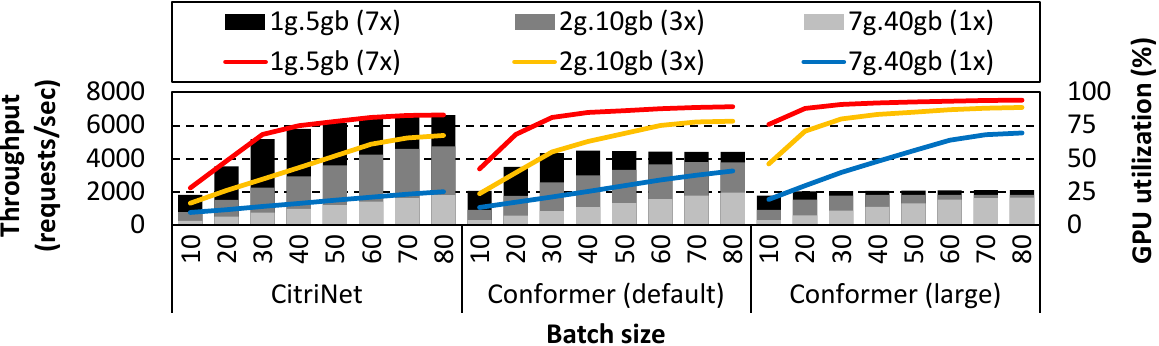}}
\vspace{-0.6em}
\caption{(Bar chart) Model execution throughput and (line chart) its GPU utilization when preprocessing is disabled. The x-axis shows the input batch size executed by a single vGPU.}
\label{fig:char_inf_only_throughput}
\end{figure}

\subsection{MIG Effectiveness in Improving GPU Utilization}
\label{sect:char_inf_only}

We first evaluate how effective MIG is in improving the ``chip-wide'' aggregate throughput, which is one of the most important motivations behind the design of MIG, i.e., addressing the GPU underutilization problem when a small AI model is serviced on top of a training-purposed large GPU. To separate out the effect of CPU-side data preprocessing on the GPU-side model execution, we manually disable the execution of the data preprocessing stage and feed the GPU with a sufficient amount of preprocessed input batches. We then measure the model execution stage's GPU utilization (\fig{fig:char_inf_only_throughput}, right axis) and the aggregate throughput  (\fig{fig:char_inf_only_throughput}, left axis) while the input batch size is fixed at a value $N$ (x-axis).

Across all MIG configurations and all AI workloads, the GPU utilization monotonically increases as the input batch size is increased. However, the rate at which GPU utilization increases is noticeably higher with the fine-grained \migsmall. This is because the compute resources of the small vGPUs in \migsmall can be utilized effectively even at small batch sizes, rendering its chip-wide GPU utilization also become high. Thanks to such high GPU utilization, the aggregate throughput a \migsmall system can achieve is significantly higher than \miglarge, even with much smaller batch sizes. 

Overall, our characterization confirms that MIG partitioned into multiple fine-grained vGPUs (\migsmall) can help improve the deployment cost of AI models by significantly improving GPU's resource utilization.

\subsection{MIG's Effect on Batching System}
\label{sect:char_batching}

\begin{figure}[t!] \centering
\subfloat[Computer vision]{\includegraphics[width=0.48\textwidth]{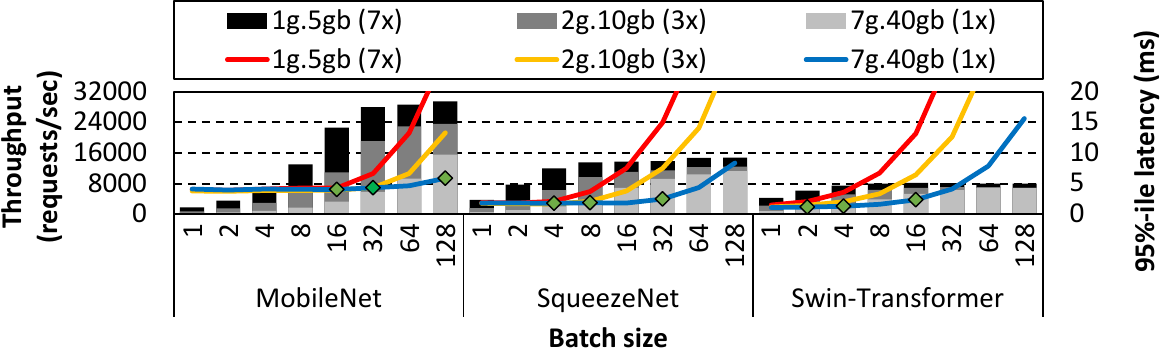}}\\
\vspace{-0.4em}
\subfloat[Audio processing]{\includegraphics[width=0.48\textwidth]{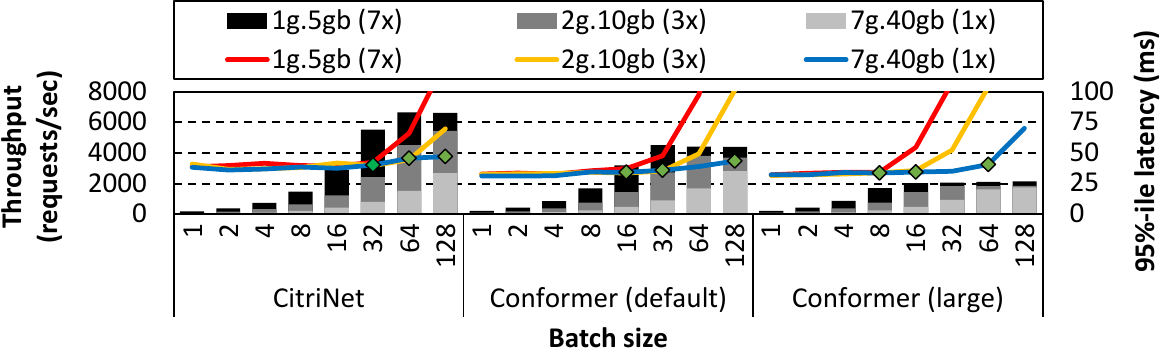}}
\vspace{-0.6em}
\caption{(Bar chart) Model execution throughput and (line chart) its $95\%$-ile latency when preprocessing is disabled.
The x-axis (log-scale) shows the input batch size executed by a single vGPU. 
}
\label{fig:char_inf_only_tail_latency}
\vspace{-1em}
\end{figure}

In real-world AI inference servers, input traffic patterns are constantly changing with varying traffic intensities, so designing an efficient batching system that balances throughput and tail latency becomes critical. In \fig{fig:char_inf_only_tail_latency}, we show how the model execution stage's throughput (left axis) and its tail latency (right axis) change as input batch size increases (x-axis). Across all MIG configurations and all models, once the aggregate throughput reaches a plateau, tail latency spikes up rapidly even with a very small increase in batch size.  If the GPU already reached close to its maximum possible throughput, further increase in the batch size directly translates into a proportional increase in execution time (i.e., tail latency) with only incremental improvements in throughput. We refer to this tipping point as ``the maximum batch size at the knee of the tail latency curve'' (\batchKnee, denoted as green diamond markers in \fig{fig:char_inf_only_tail_latency}).  Given the property of \batchKnee, we can see that setting the \emph{maximum batch size} (\batchMax, i.e., the largest possible batch size the batching system will try to construct) as the \batchKnee value is \emph{optimal} because having \batchMax larger than \batchKnee provides practically no gain in throughput while only aggravating tail latency.

\begin{figure}[t!] \centering
\includegraphics[width=0.47\textwidth]{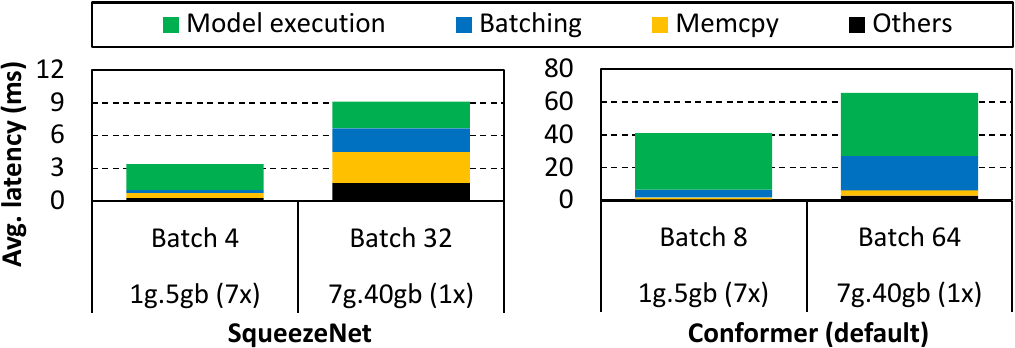}
\vspace{-0.6em}
\caption{Breakdown of average latency when \migsmall and \miglarge are
configured with the appropriate \batchMax value to sustain the same
inference throughput. Data preprocessing stage is disabled in this experiment.}
\label{fig:char_latency_breakdown}
\end{figure}

As the batch size is increased, MIG partitioned into many small vGPUs (\migsmall) experiences a more rapid increase in aggregate throughput than \migmed and \miglarge. This is because \batchKnee value for \migsmall is smaller than that of \migmed and \miglarge. For example, \migsmall has  \batchKnee of $16$/$4$/$2$  whereas \miglarge has a value of $128$/$32$/$16$ for MobileNet/SqueezeNet/Swin-Transformer. Because \migsmall servers can reach high throughput even with small batches, an interesting property of \migsmall is that \emph{the batching system can spend less time waiting for inputs to construct a batch} vs. \miglarge. \fig{fig:char_latency_breakdown} breaks down average latency when both \migsmall and \miglarge are each configured with the proper \batchMax value that provides the \emph{same} end-to-end inference throughput. As depicted, \migsmall has a smaller \batchMax value than \miglarge so it spends less time constructing batched inputs  (the blue-colored ``Batching'').  In \sect{sect:proposed}, we utilize these properties to design a dynamic batching system for MIG.

\begin{figure}[t!] \centering
\subfloat[Computer vision]{\includegraphics[width=0.46\textwidth]{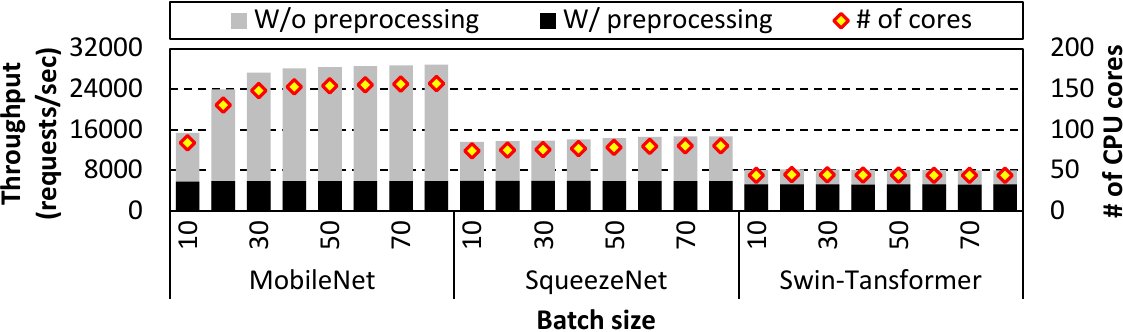}}\\
\vspace{-0.4em}
\subfloat[Audio processing]{\includegraphics[width=0.46\textwidth]{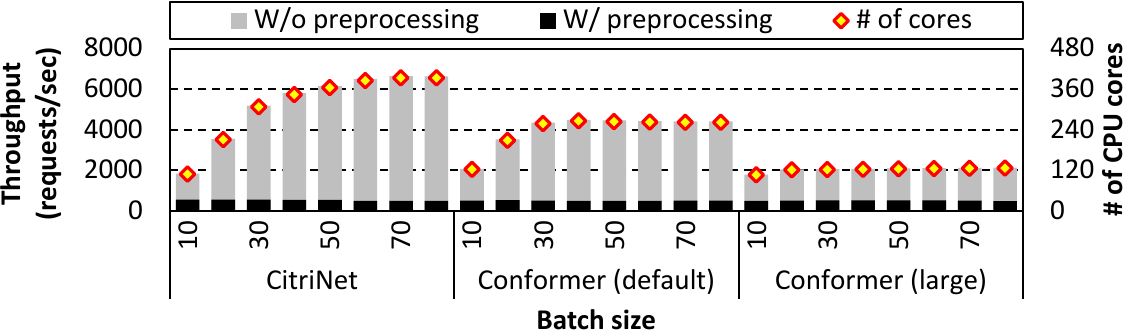}}
\vspace{-0.6em}
\caption{(Left axis) Inference throughput with (black) and without (grey) the data preprocessing stage. (Right axis) The minimum number of CPU cores required for just preprocessing to sustain a single A100 GPU's model execution throughput (i.e., matching the throughput of gray bars). Experiments are conducted over a server containing $32$ CPU cores with \migsmall MIG configuration.}
\label{fig:char_e2e_throughput_without_preproc}
\end{figure}

\subsection{Effect of Data Preprocessing on MIG  Throughput}
\label{sect:char_data_preproc_effect}

We now turn our attention to data preprocessing, evaluating its effect on end-to-end performance. The left axis in \fig{fig:char_e2e_throughput_without_preproc} compares the end-to-end inference throughput with and without data preprocessing. That is, the performance \emph{without} the data preprocessing stage in \fig{fig:char_e2e_throughput_without_preproc} (gray bars) is identical to the performance reported in \fig{fig:char_inf_only_throughput}, whereas the performance \emph{with} data preprocessing is collected by executing all three stages in \fig{fig:inference_pipeline} end-to-end.

As depicted, the overall performance experiences a significant slowdown when the data preprocessing stage is activated, causing a $75.6\%$ drop in throughput. With the \migsmall design point, a total of seven AI inference servers can concurrently process the service queries and cause a proportional increase in data preprocessing demand, i.e., $7\times$ higher than \miglarge. Such performance drop is due to the limited amount of CPU-side resources available for data preprocessing. Specifically, several key data preprocessing operations require abundant CPU cores and memory bandwidth in order to transform the raw input into AI model-specific formats. Unfortunately, we observe that the compute throughput and memory bandwidth in current CPUs are not sufficiently high enough to sustain the data preprocessing requirements of multiple inference servers deployed within MIG.  In the right axis of \fig{fig:char_e2e_throughput_without_preproc}, we show the minimum number of CPU cores required just for preprocessing to fully unlock the maximum possible inference throughput (i.e., one that matches the AI model execution stage's throughput, the gray bars in \fig{fig:char_e2e_throughput_without_preproc}). It is worth emphasizing that the host CPU is already busy handling various other critical tasks (e.g., load-balancing input traffic, GPU kernel launching, etc) so executing the data preprocessing stage puts high pressure on the already heavily burdened host CPU. Consider CitriNet which requires a staggering 393 preprocessing CPU cores (refer to \sect{sect:methodology}) just to sustain the serving of a \emph{single} A100 GPU partitioned as \migsmall, highlighting the significant bottleneck data preprocessing incurs.

\begin{figure}[t!] \centering
\subfloat[Computer vision]{\includegraphics[width=0.47\textwidth]{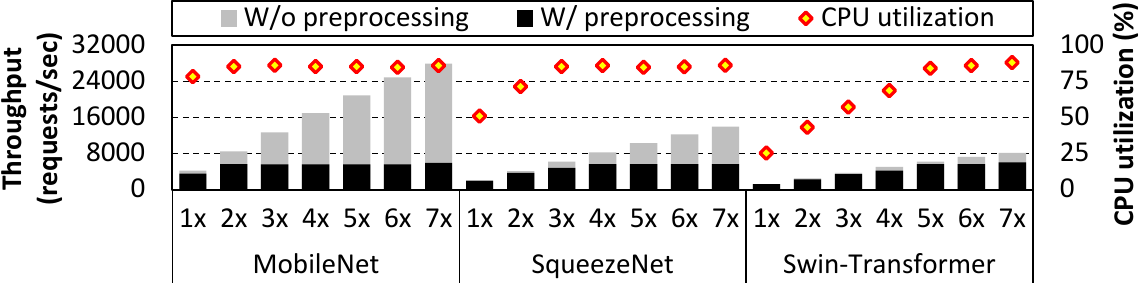}}\\
\vspace{-0.4em}
\subfloat[Audio processing]{\includegraphics[width=0.47\textwidth]{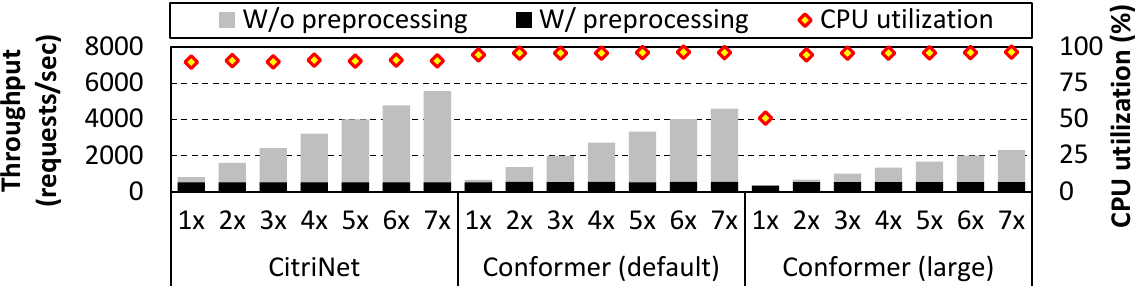}}
\vspace{-0.4em}
\caption{Throughput (left axis) and CPU utilization (right axis) as a function of the number of inference servers activated, each server executing on a vGPU within a \migsmall MIG configuration.}
\label{fig:char_analysis_sm_utility_and_throughput_scalability}
\end{figure}

Such phenomenon is better highlighted in \fig{fig:char_analysis_sm_utility_and_throughput_scalability} which shows how the data preprocessing throughput scales as a function of the number of inference servers deployed within MIG. In this experiment, the MIG is first configured as \migsmall and is instantiated with seven inference servers initially.  We then limit the number of inference servers activated, from one to seven, and measure the data preprocessing throughput (left axis) as well as the corresponding CPU utilization  (right axis). As shown, for all models we evaluate, the CPU utilization saturates around $90\%$ with only a small number of inference servers activated and fails to further scale up throughput beyond this point.  This implies that MIG's inference servers will starve for preprocessed input data it can consume, leaving the GPU-side model execution stage idle.

\section{\proposed Architecture}
\label{sect:proposed}

\subsection{High-Level Overview}
\label{sect:epb_overview}

\begin{figure}[t!] \centering
  \includegraphics[width=0.47\textwidth]{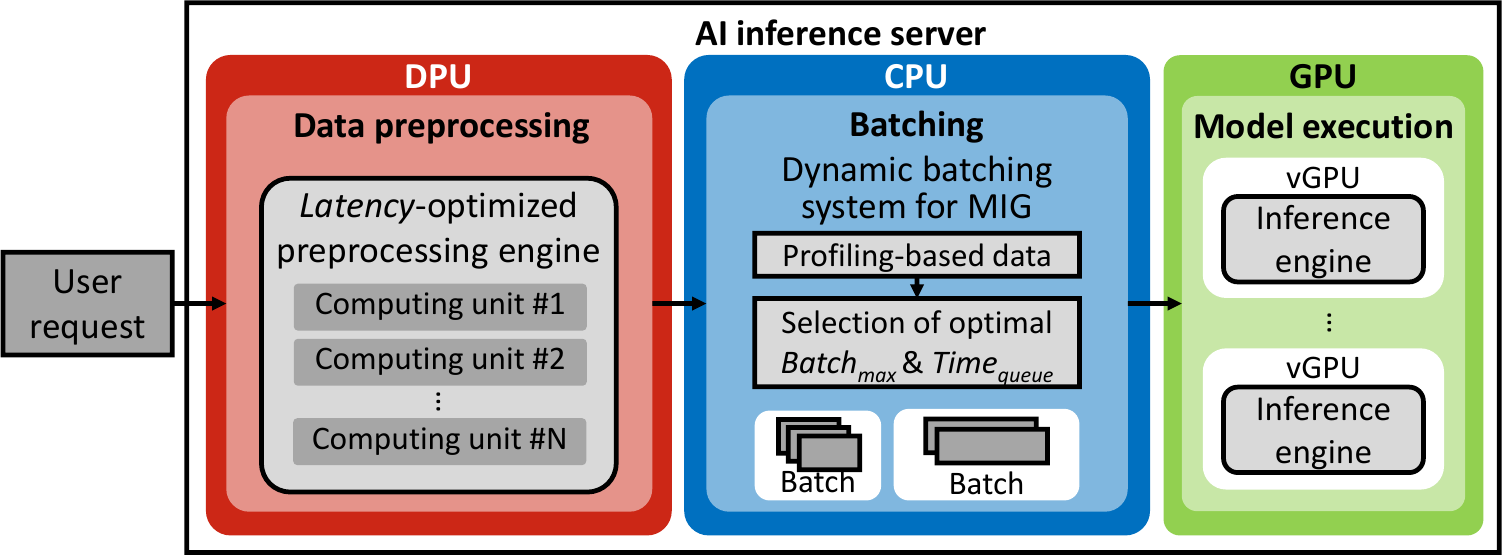} 
  \vspace{-0.4em}
  \caption{High-level overview of \proposed.}
  \label{fig:epb_overview}
\end{figure}

\fig{fig:epb_overview} provides a high-level overview of our \proposed system.
Our DPU unlocks the full potential of MIG by fundamentally addressing its data
preprocessing bottlenecks with domain-specific acceleration. Concretely, all
inference requests routed to the MIG inference server are completely offloaded
to our DPU, which is integrated at the PCIe bus as the host CPU's
\emph{co-processor}, providing substantial improvements in preprocessing
throughput.  Once the MIG inference server is unleashed from its preprocessing
overheads, our software system utilizes our dynamic batching system
for high-performance AI inference.

\subsection{DPU Architecture for MIG Data Preprocessing}
\label{sect:dpu_arch}
\label{page:dpu_design_philosophy}

{\bf Design objective.} We architect our DPU with the following two design
objectives:  1) \emph{design flexibility} and 2) \emph{latency-centric}
microarchitecture.  Because of the diversity of AI inference workloads, we
utilize an FPGA as our design substrate to flexibly handle various
application-specific preprocessing operations while also reaping out the
benefits of domain-specific acceleration.  Furthermore, the design of our DPU
microarchitecture is optimized for minimizing the latency to preprocess a batch
with just a \emph{single} input request while also maximizing aggregate
throughput.

{\bf Flexible design based on reconfigurable hardware.} In this paper, we focus on the DPU implementation for computer vision and audio processing AI models. While multiple AI models have evolved, preprocessing methods have remained relatively unchanged due to the limited diversity of input modalities (image/audio/text). By offering distinct implementations for both image and audio, we ensure that the majority of preprocessing computations can be efficiently offloaded with only incremental adjustments to the current implementation, while disregarding text-based modalities, which require minimal preprocessing (e.g., tokenization). Additionally, the implementation leverages the user-friendly HLS language alongside reconfigurable hardware to maximize deployment flexibility.

{\bf Motivation for single-input batch optimization.} Hardware accelerators for
AI are typically optimized for batches with multiple input requests as it helps
maximize parallelism and overall throughput. In contrast, our DPU microarchitecture is
optimized for \emph{single}-input batches for several reasons.

\begin{enumerate}

\item As we characterized in \sect{sect:char_batching}, the performance of a
MIG inference server is highly sensitive to the \batchMax value, the optimal
value of which is determined by the AI model architecture and the MIG
partitioning granularity, i.e., the vGPU size.  Optimizing the preprocessing
stage for fast single-input preprocessing enables requests to be preprocessed
immediately upon arrival to the inference server, providing extra latency
budget for the next batching stage (\fig{fig:inference_pipeline}) to identify
the optimal batching strategy in accordance to \batchMax (the maximum batch size) and
\batchQueue (the maximum queueing delay for batching).

\begin{sloppypar}\tolerance 900
\item More crucially, the relatively scarce compute/memory resources available
in \migsmall's individual vGPUs render its optimal \batchMax value to become much
smaller than \miglarge's \batchMax value (e.g., Swin-Transformer's \batchMax value is only $2$
under \migsmall).  Therefore, having the input requests be preprocessed in
the most finest granularity (i.e., single-input request) provides the most
flexibility to the subsequent batching stage to construct any batch size that best
fulfills the model execution stage's need.
\end{sloppypar}

\end{enumerate}

\begin{figure}[t!] \centering
    \vspace{-1em}
    \subfloat[Computer vision]{\includegraphics[width=0.47\textwidth]{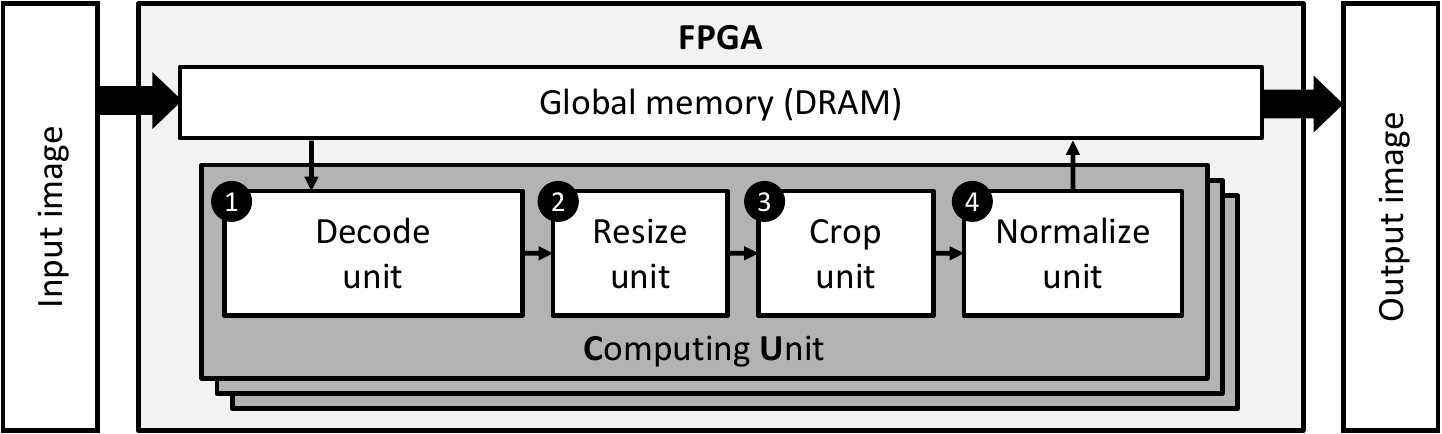}}\\
    \vspace{-0.4em}
    \subfloat[Audio processing]{\includegraphics[width=0.47\textwidth]{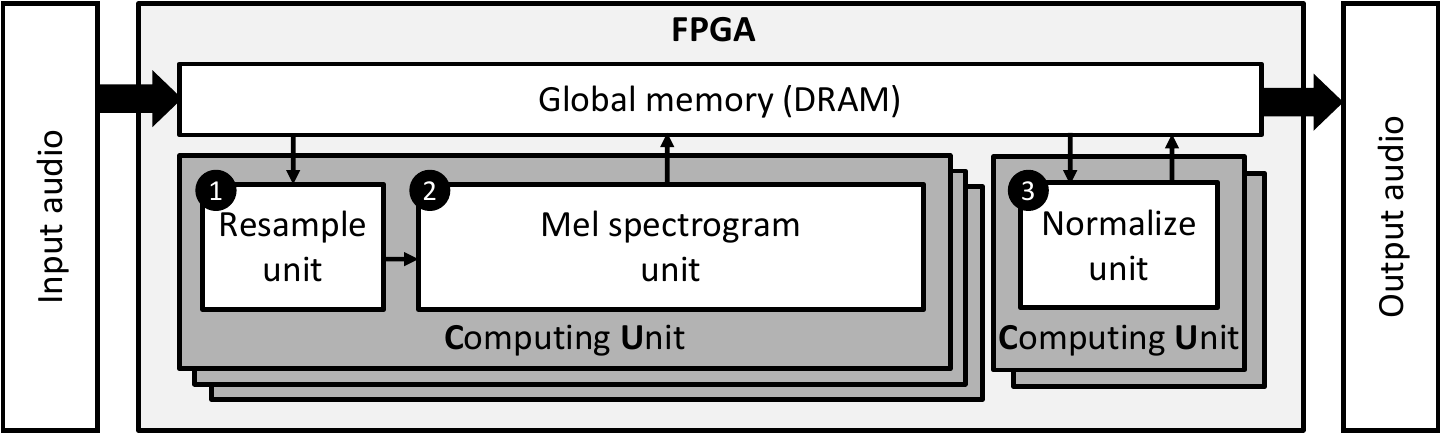}}\\
    \vspace{-0.4em}
    \caption{DPU microarchitecture for preprocessing (a) computer vision and (b) audio processing. Each functional unit within the CU handles its own preprocessing operation defined in \fig{fig:data_preprocessing}.}
    \label{fig:dpu_arch}
\end{figure}

{\bf DPU microarchitecture.} \fig{fig:dpu_arch} shows our DPU microarchitecture targeting computer vision and audio processing workloads, each of which contains multiple {\bf C}omputing {\bf U}nits (CUs) for preprocessing. A CU is the smallest granularity in which the host CPU controls and communicates with the FPGA to 1) transfer input/output data and 2) command the FPGA to execute
preprocessing operations.  A given CU contains multiple functional units, each of which is responsible for a specific stage of preprocessing (e.g., the ``Decode'' unit in a CU handles JPEG decoding operation in \fig{fig:data_preprocessing}(a))\footnote{ For
clarity of explanation, we use the terminologies defined in Xilinx's HLS programming guide~\cite{hls_xilinx}. The host CPU communicates with the CU using command queues and buffers created by Xilinx OpenCL extensions~\cite{opencl,xilinx_opencl}.  Different functional units within a CU are designed to directly forward its input/output data via on-chip buffers (without communicating it over off-chip DRAM) in accordance with its \emph{stream} abstraction~\cite{hls_xilinx} using FIFO queues.}. Each functional unit is carefully designed to maximally reap out data-level parallelism inherent in the single input request leveraging the APIs on Xilinx Vitis DSP and Vision Library\cite{vitis_libraries}. As mentioned previously, our CU is designed to minimize latency to process a single-input batch request. Despite such \emph{latency}-optimized DPU design, we also seek to
maximize aggregate \emph{throughput} by employing multiple CUs in parallel (\mbox{\fig{fig:epb_overview}}). As such, our DPU can leverage \emph{request-level parallelism} to concurrently preprocess multiple single-input batches.

\begin{figure}[t!] \centering
    \captionsetup[subfloat]{captionskip=-0.3em}
    \subfloat[]{\includegraphics[width=0.47\textwidth]{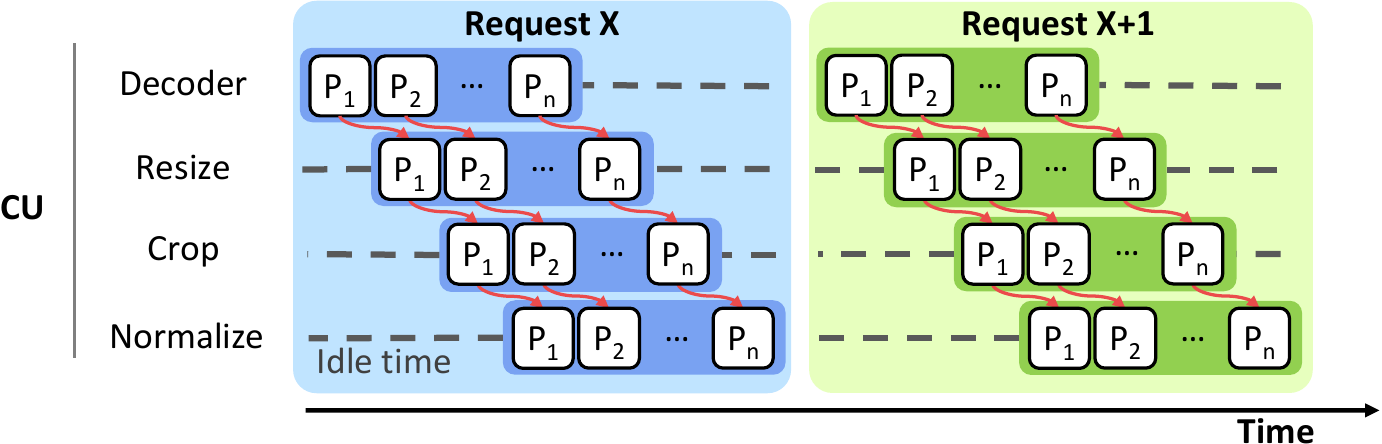}}\\
    \vspace{-0.4em}
    \subfloat[]{\includegraphics[width=0.47\textwidth]{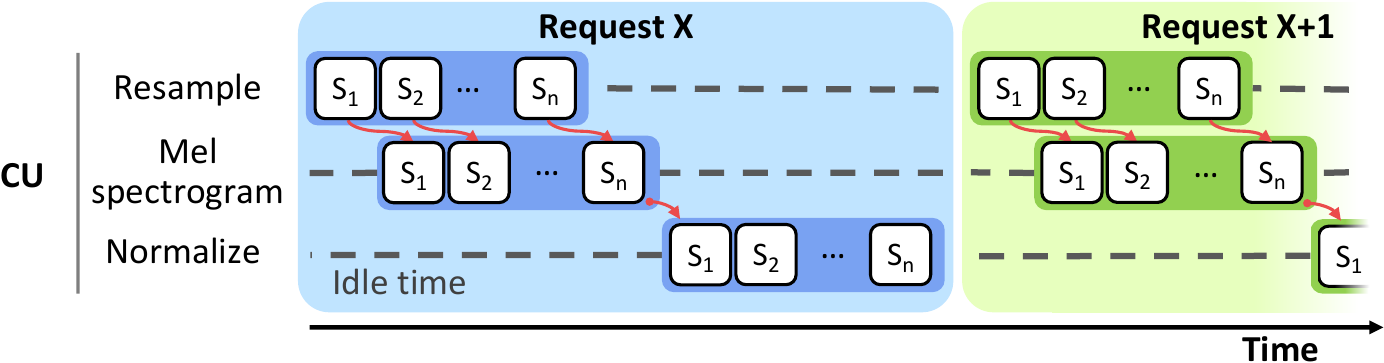}}\\
    \captionsetup[subfloat]{captionskip=0.1em}
    \vspace{-0.4em}
    \subfloat[]{\includegraphics[width=0.47\textwidth]{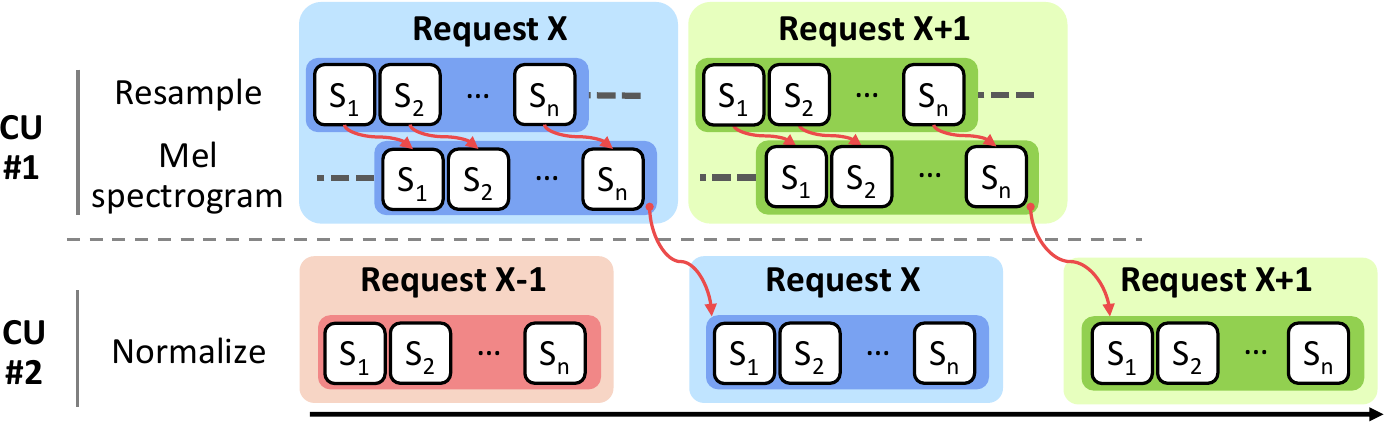}}\\
    \vspace{-0.4em}
    \caption{Execution timeline how the two requests ($X$ and $X$+1) are handled by our DPU in (a) computer vision and (b,c) audio processing workloads. The timeline in (b) assumes a DPU design that integrates \emph{all} of the functional units of audio processing within a single CU, whereas (c) assumes our proposed DPU design in \fig{fig:dpu_arch}(b) that utilizes two separate CU types for better resource utilization. P$_{i}$ and S$_{i}$ each refer to the group of image pixels and audio samples preprocessed by our functional units, respectively.}
    \label{fig:dpu_timeline}
\end{figure}

It is worth emphasizing that a DPU microarchitecture that singlehandedly focuses on minimizing single-input preprocessing latency, without considering the DPU's overall utilization, leads to sub-optimal system-wide throughput. Consider our DPU design for computer vision in \fig{fig:dpu_arch}(a) and how it handles the two single-input batches in \fig{fig:dpu_timeline}(a). In the preprocessing algorithm for computer vision, different operations within data preprocessing have a simple, sequential inter-operation data dependency because the previous operation's output is only used as the next operation's input (\fig{fig:dpu_timeline}(a)). This type of sequential dataflow enables a single CU to integrate all types of functional units while still being able to fully utilize all of these units via pipelined execution (\fig{fig:dpu_timeline}(a)). This is in stark contrast to the preprocessing algorithm for audio processing which has a relatively complex dataflow. In audio preprocessing, the ``Normalize'' unit executes the following three sub-operations in sequence: 1) calculating the mean value across all the input audio samples, 2) determining the variance of all samples using the mean value calculated in the previous step\#1, and 3) normalizing each sample data based on the mean and variance values derived previously. As such, the ``Normalize'' unit cannot initiate its preprocessing operation until \emph{all} the input samples have been processed by the previous ``Resample'' and ``Mel spectrogram'' units. Putting it differently, even when two single-input batch requests are available for scheduling, a CU design that integrates all functional units of audio preprocessing (similar to how a CU is designed for computer vision, \fig{fig:dpu_arch}(a)) can only start preprocessing the second request when the first request is fully preprocessed (\fig{fig:dpu_timeline}(b)). Our latency-optimized DPU tackles such challenge by designing two separate CU types (i.e., CU for ``Resample and Mel spectrogram'' and CU for ``Normalize'', \fig{fig:dpu_arch}(b)) which our software runtime system utilizes for fine-grained scheduling of single-input batches. Such design point enables each single-input request to be processed with minimal latency while making sure the aggregate DPU-wide throughput is maximized via better utilization of our functional units (\fig{fig:dpu_timeline}(c)).

\phantomsection\label{page:PCIe_bandwidth}
{\bf Implication of adding DPU to the system.} DPU is integrated into the system as a PCIe-attached I/O device. Below we discuss the potential overheads it may introduce.

\textit{Latency overhead.} 
In PREBA, the preprocessed data (which is generated by the DPU) is first forwarded back to the CPU, which is then sent to the GPU for model execution, i.e., DPU$\rightarrow$CPU$\rightarrow$GPU. One might be concerned that adding such an extra round trip latency can potentially nullify the performance benefits our FPGA-accelerated preprocessing provides. However, this extra PCIe latency is measured in the order of tens of microseconds, whereas an end-to-end inference request typically takes a few to tens of milliseconds. Therefore, the additional latency incurred by having our DPU integrated as a separate
I/O device is negligible. 

\textit{PCIe bandwidth.} 
Our DPU is integrated into the PCIe root complex via a PCIe switch, functioning as an additional I/O device from the host CPU's perspective. Having an additional I/O device in the system can therefore potentially pressurize the PCIe communication bandwidth. However, the maximum usage of PCIe bandwidth for image and audio processing workloads incurring frequent CPU$\leftrightarrow$DPU data transfers (e.g., MobileNet and CitriNet) is measured at 6.13 GB/s and 0.9 GB/s, respectively. These numbers are significantly lower than the 32 GB/s communication bandwidth available with PCIe(gen4) specification (and more so with the now prevalent PCIe(gen5)), rendering DPU's integration at PCIe to cause minimal interference to other I/O devices. However, in scenarios where multiple DPUs and GPUs are integrated within the same
PCIe root complex, it is possible that our PREBA system can be bottlenecked by PCIe bandwidth constraints. In those circumstances, implementing P2P data movement between DPU$\leftrightarrow$GPU across a PCIe switch and leveraging the switch's bandwidth for inter-device communication can significantly alleviate any potential communication bottleneck \mbox{\cite{park2020trainbox}}.

\label{page:TCO_1}
\textit{Cost.}  
Because our DPU is implemented using a separate FPGA card, it introduces extra cost and impacts the TCO of maintaining an inference server. In \sect{sect:eval_tco}, we quantitatively evaluate PREBA's cost-efficiency. 

\begin{figure}[t!] \centering
    \includegraphics[width=0.47\textwidth]{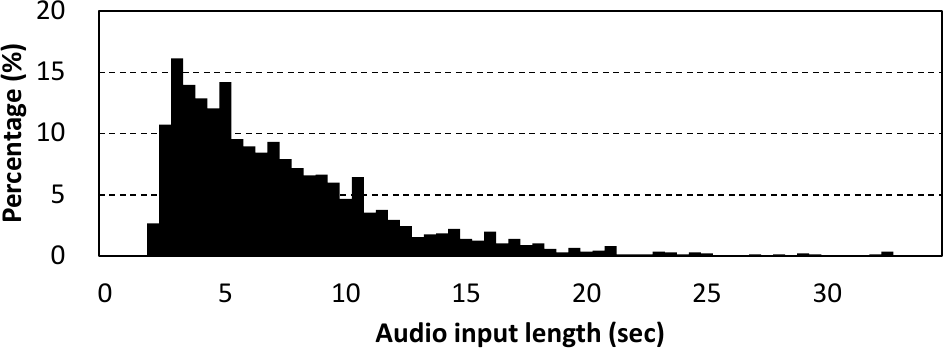}
    \vspace{-0.4em}
    \caption{Histogram of Librispeech's audio input lengths~\cite{librispeech}}
    \label{fig:librispeech_dataset}
\end{figure}

\subsection{Dynamic Batching System for MIG}
\label{sect:batching}

Thanks to our DPU's single-input batch-optimized design philosophy, our batching system can freely construct any input batch size that fulfills the requirements of the model execution stage.  An efficient batching system requires the following two hyperparameters to be carefully chosen for optimal batching: \batchMax (the largest batch size the batching system will try to construct) and \batchQueue (the maximum time period the batching system will have input requests wait in a batching queue to construct up to \batchMax-sized input batch).  As discussed in \sect{sect:char_batching}, the optimal value of \batchMax is when it is set to the \batchKnee point. This is because constructing a batch that is larger than \batchKnee provides no benefits in improving throughput while significantly harming tail latency.  Note that the \batchKnee value is determined based on several factors: 1) the MIG ``hardware'' configuration inference is undertaken (\migsmall vs.  \miglarge), 2) the AI ``model'' subject for inference (MobileNet vs. Swin-Transformer), and 3) the size of the model ``input'' (a fixed-size (224$\times$224$\times$3) input image vs. a variable-length audio input sample). We already discussed how the first two factors (``hardware'' and ``model'') affect the \batchKnee point in \fig{fig:char_inf_only_tail_latency}, so let us focus our discussion on the third factor, the model ``input'' size. 

\begin{figure}[t!] \centering
    \vspace{-1.5em}
    \subfloat[\migsmall]{\includegraphics[width=0.24\textwidth]{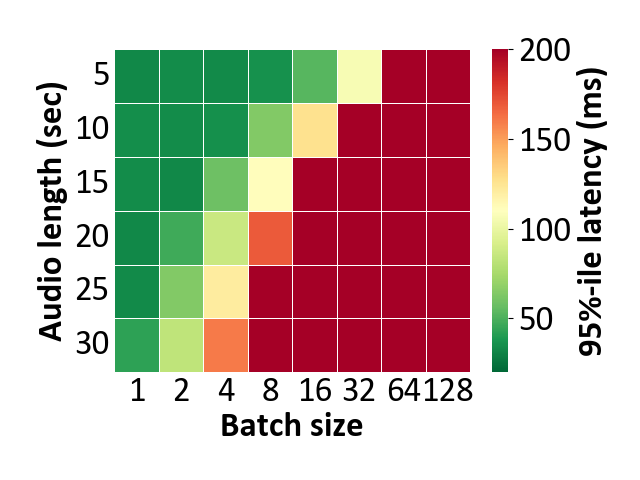}}
    \subfloat[\miglarge]{\includegraphics[width=0.24\textwidth]{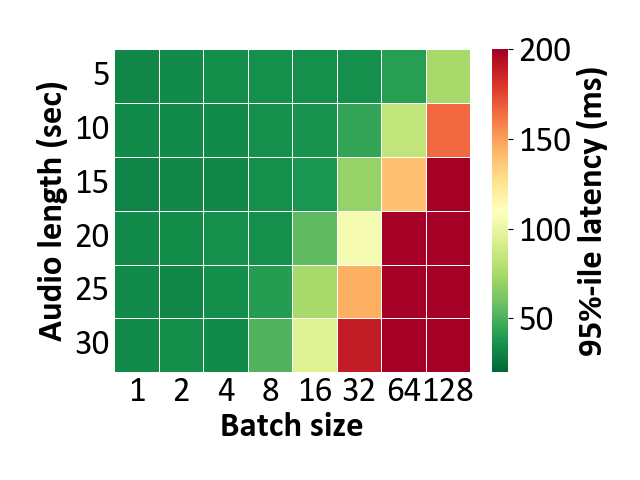}}\\
    \vspace{-0.6em}
    \caption{Effect of input batch size (x-axis) and audio input length (y-axis) on tail latency (different colors in the heat map, 0$-$200 ms) over (a) \migsmall and (b) \miglarge when executing Conformer(default). The \batchKnee point corresponds to the point where the heat map color rapidly transitions from green to yellow.}
    \label{fig:impact_of_batch_and_length_on_latency_audio}
\end{figure}

\begin{figure}[t!] \centering
    \vspace{-1.2em}
    \includegraphics[width=0.47\textwidth]{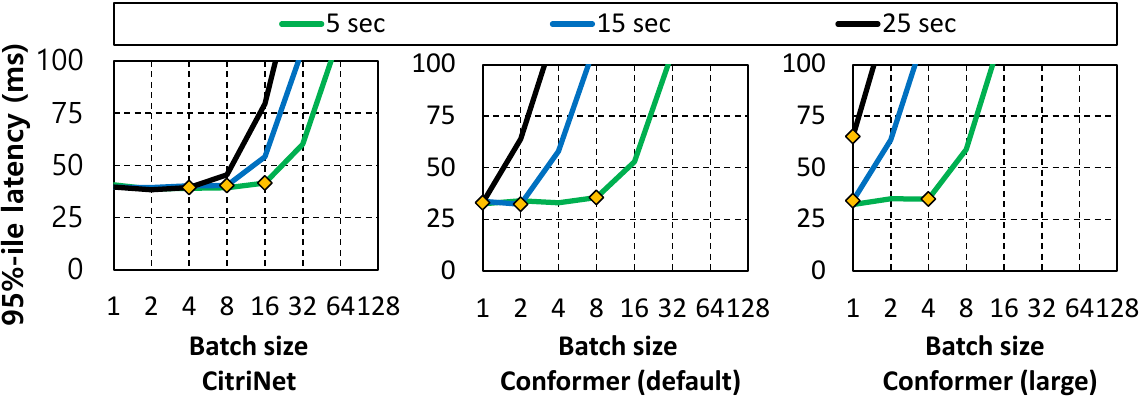}
    \vspace{-0.4em}
    \caption{Tail latency of the CitriNet and two Conformer models over \migsmall when the audio input length changes among 5/15/25 seconds. Note that the tail latency value at \batchKnee (denoted as diamond markers) is almost constant around $35$ ms (i.e., \tailKnee), regardless of the audio input length.}
    \label{fig:tail_latency_conformer}
\end{figure}

{\bf Effect of variable-length inputs on batching.} Unlike data
preprocessing for computer vision which always generates a fixed-size
(224$\times$224$\times$3) image for the model execution stage, the audio input
length for audio processing can vary significantly
(\fig{fig:librispeech_dataset}). As such, the \batchKnee point of an audio
processing model can vary depending on the audio input length which
we visualize in \fig{fig:impact_of_batch_and_length_on_latency_audio} and
\fig{fig:tail_latency_conformer}. While
different audio input lengths lead to different \batchKnee points, the tail
latency value at \batchKnee is almost constant at around 35 ms, regardless of the
audio input length. In the remainder of this section, we refer to this tail
latency value at \batchKnee as \tailKnee.  Below we describe how \proposed's batching system
utilizes the aforementioned properties of \batchKnee and \tailKnee to
systematically determine the optimal \batchMax and \batchQueue for MIG.

{\bf Profiling-based estimation of Batch$_{max}$.} \proposed conducts an offline profiling of the throughput vs. tail latency curve as a function of batch size and input size for the target AI model served on top of the MIG configuration of interest. The cost of conducting this one-time profiling is very low (several minutes to collect all results) and more importantly, such overhead is amortized over the several millions to billions of user queries serviced by the MIG inference server. Because the input size is fixed for computer vision while variable-length for audio processing, the profiled data will be in the form of \fig{fig:char_inf_only_tail_latency} for computer vision and be
\fig{fig:impact_of_batch_and_length_on_latency_audio}/\fig{fig:tail_latency_conformer} for audio processing. The \batchKnee points can then be derived by utilizing these profiled data. For instance, only a single \batchKnee point exists for a computer vision model (\fig{fig:char_inf_only_tail_latency}) so we configure its \batchMax identical to this \batchKnee point. As for audio processing, multiple \batchKnee points exist per model (\fig{fig:tail_latency_conformer}), one per each audio input length, meaning there can also be multiple \batchMax values that are optimal for each audio input length. 

\begin{figure}[t!] \centering
    \includegraphics[width=0.47\textwidth]{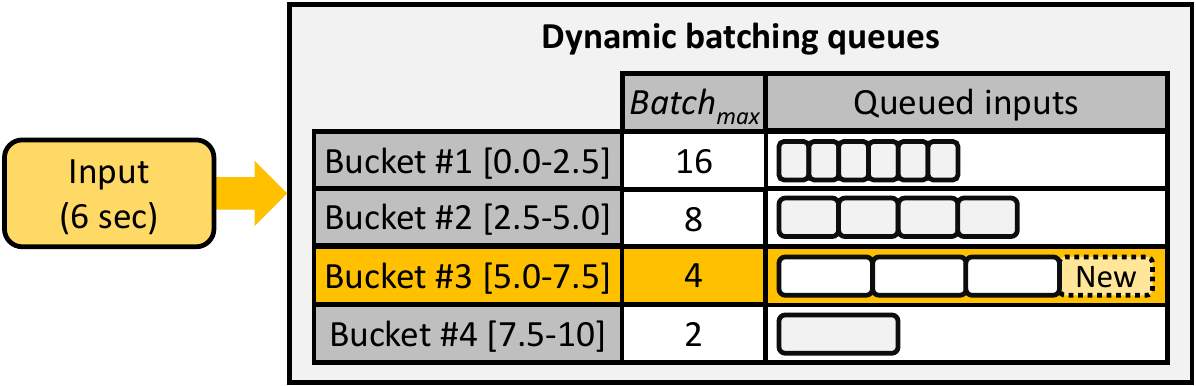}
    \caption{\proposed's dynamic batching system for audio processing workloads. Since the input request has an audio input length of 6 seconds, it falls under the third bucket, which allows the batching unit to coalesce the four input requests into a single batch of size $4$. Example assumes the bucket window is 2.5 seconds long.}
    \label{fig:dynamic_batching}
\end{figure}

\fig{fig:dynamic_batching} illustrates \proposed's \emph{dynamic} batching
system that considers the variable-length nature of audio inputs and
the corresponding \batchKnee and \batchMax values for optimal batching.  In our
proposed batching system, the audio input lengths are bucketized into multiple
non-overlapping windows of $2.5$ seconds (e.g., [0.0$-$2.5 sec], [2.5-5.0 sec],
		$\ldots$). Each bucket is allocated with a dedicated \emph{batching queue}
that buffers all audio input requests whose length falls under that bucket.
Therefore, a MIG inference server maintains a total of $N$ batching queues
($N$: total number of buckets).  Because each batching queue is configured with
its own optimal \batchMax value, one which is set identically to the \batchKnee
point that was derived through profiling, \proposed can dynamically construct
the appropriate batch size that meets the properties of variable-length audio
inputs.  Putting everything together, once an input request is routed to our
\proposed inference server, the audio input length is examined and is
bucketized to be forwarded to its corresponding batching queue. Our dynamic
batching system then seeks to accumulate enough audio inputs to construct a
batch size that is up to \batchMax.

{\bf Analytical model based estimation for Time$_{queue}$.} The purpose of
\batchQueue is to regulate the time a batching system spends trying to form a
batch. When the input traffic intensity is low (e.g., only a handful of requests are routed to the server), spending too much time waiting to form an input batch of size \batchMax can be wasteful. This is because the time spent trying to form that big of a batch size may incur too much latency and leave several vGPUs idle.

As such, \batchQueue should be chosen to minimize the idle time of GPU while making sure the batching system has a sufficient window of time to construct (up to \batchMax-sized) input batches. We previously established that \batchMax is the optimal batch size an inference server can employ in order to maximize throughput while maintaining tail latency within SLA bounds. Because our batching system does not construct batches whose size is larger than \batchMax, we can
safely assume that the GPU's model execution time of \emph{any} given input batch size will always be shorter than \tailKnee (i.e., tail latency at \batchKnee).  Consequently, for an inference server with a single vGPU (\miglarge), setting the \batchQueue value identical to \tailKnee is optimal because the batching system only spends as much time batching the inputs that match the time GPU spends executing an existing input batch.  As our \migsmall inference server contains seven vGPUs, the \batchQueue value should therefore be adjusted to make sure all
seven vGPUs always have at least a single input batch to consume. In \proposed, we set the \batchQueue time as the (\tailKnee of a single vGPU executing the target AI	model using \batchMax-sized input)/(total number of vGPUs, seven in \migsmall). This guarantees that the batching system at least generates an average of seven new batched inputs while the seven vGPUs concurrently execute each of its existing input batch for an average execution time of \tailKnee.

\proposed divides variable-length audio inputs into multiple buckets based on their lengths. For each bucket, PREBA checks to ensure that there are no requests waiting longer than the \batchQueue time. However, because requests are split across multiple queues, it is possible that GPU utilization can become low when insufficient number of requests are batched in each queue. To alleviate such inefficiency, if the constructed batch size does not reach the corresponding \batchMax for that bucket, PREBA tries to merge requests from adjacent buckets to maximize GPU utilization. When merging requests from different buckets, our batching system ensures that the batch size does not exceed the \batchMax of the longest input within the batch.

\section{Methodology}
\label{sect:methodology}

\textbf{Benchmarks.} We study a total of six AI workloads from computer vision
(MobileNet (v3 small)~\cite{mobilenet}, SqueezeNet (v1.1)~\cite{squeezenet},
and Swin-Transformer~\cite{swin}) and audio processing (two different model sizes of
Conformer~\cite{conformer} and CitriNet~\cite{citrinet}). We chose these models as they exhibit different computational intensities, allowing us to demonstrate the effectiveness of \proposed over different model characteristics.  The model parameters of all of our computer vision models are from TorchHub~\cite{torch_hub} while those for audio processing are from NVIDIA NeMo~\cite{nvidia_nemo}.

\textbf{Input query modeling.} Following prior work~\cite{clipper,mark}, we assume that each service query arrives to the inference server as a single-input request. In terms of
query arrival rates, we follow recommendations from the MLPerf inference benchmark~\cite{mlperf_inf}
which assumes a Poisson distribution for modeling the rate at which a new query arrives to the 
server. As for the input data, we used the ILSVRC-2012 dataset~\cite{ilsvrc2012}
for computer vision workloads and LibriSpeech dataset~\cite{librispeech} for audio processing workloads.

\textbf{Hardware.} 
Our server platform contains an AMD EPYC 7502 CPU, which contains 32 physical CPU cores with $256$ GB of DRAM. We utilize a single NVIDIA A100 GPU~\cite{a100}
to design our MIG inference server which communicates with the CPU over PCIe. \proposed's
DPU is implemented on Xilinx's U55C FPGA~\cite{u55c} using Vitis HLS
2022.2~\cite{vitis}. The DPU communicates with the CPU over PCIe. 
\tab{tab:audio_fpga} summarizes our DPU's FPGA resource utilization.

\textbf{Software.} 
NVIDIA Triton Inference Server (TRTIS)~\cite{trtis} currently does not support MIG (i.e., a single CUDA process
can only enumerate a single MIG device), so we design a custom AI 
inference server faithfully following the design philosophy
of TRTIS using Python 3.8.10, PyTorch 2.0.0, and CUDA 12.1.
We utilize the Xilinx Runtime library to manage \proposed's DPU. The baseline CPU-based data preprocessing stages are implemented using OpenCV~\cite{opencv} for computer vision and Librosa~\cite{librosa} for audio processing.

\renewcommand{\arraystretch}{1.2}
\begin{table}[]
\scriptsize
\begin{tabular}{|c|c|c|c|c|c|c|}
\hline
\textbf{Application}            & \textbf{Unit}                                                       & \textbf{LUT}    &\textbf{REG}    &  \textbf{BRAM}   & \textbf{URAM}   & \textbf{DSP}    \\ \hline
\hline
\multirow{5}{*}{Image} & Decode                                                     & 19.7\% & 8.6\%  & 0.7\%  & 22.5\% & 6.2\%  \\ \cline{2-7}
                       & Resize                                                     & 7.1\%  & 2.3\%  & 0.0\%  & 0.0\%  & 8.6\%  \\ \cline{2-7}
                       & Crop                                                       & 0.6\%  & 0.4\%  & 0.0\%  & 0.0\%  & 0.0\%  \\ \cline{2-7}
                       & Normalize                                                  & 13.0\% & 3.3\%  & 11.2\% & 0.0\%  & 3.0\%  \\ \cline{2-7}
                       & Total                                                      & 44.5\% & 16.5\% & 19.3\% & 22.5\% & 17.8\% \\ \hline
\hline
\multirow{4}{*}{Audio} & Resample                                                   & 0.2\%  & 0.1\%  & 1.0\%  & 0.0\%  & 0.0\%  \\ \cline{2-7}
                       & \begin{tabular}[c]{@{}c@{}}Mel \\ spectrogram\end{tabular} & 41.5\% & 24.6\% & 18.2\% & 37.5\% & 34.2\% \\ \cline{2-7}
                       & Normalize                                                  & 3.1\%  & 1.7\%  & 1.7\%  & 7.5\%  & 1.3\%  \\ \cline{2-7}
                       & Total                                                      & 45.9\% & 26.9\% & 23.3\% & 45.0\% & 35.5\% \\ \hline
\end{tabular}
\caption{FPGA resource utilization of our proposed DPU. }  
\label{tab:audio_fpga}
\vspace{-1.3em}
\end{table}

\section{Evaluation}
\label{sect:evaluation}

In this section, all MIG inference servers are assumed to be configured as \migsmall as default (we later discuss the efficacy of \proposed over different configurations as a sensitivity study).
The ``Ideal'' design refers to an oracular, upper-bound system which assumes that preprocessing does not cause any performance overhead.``Preprocessing(DPU)'' and ``Preprocessing(CPU)'' refer to our \proposed system and the baseline CPU-based preprocessing design, respectively.

\subsection{Performance}
\label{sect:eval_perf}

\begin{figure}[t!] \centering
\subfloat[Computer vision]{\includegraphics[width=0.47\textwidth]{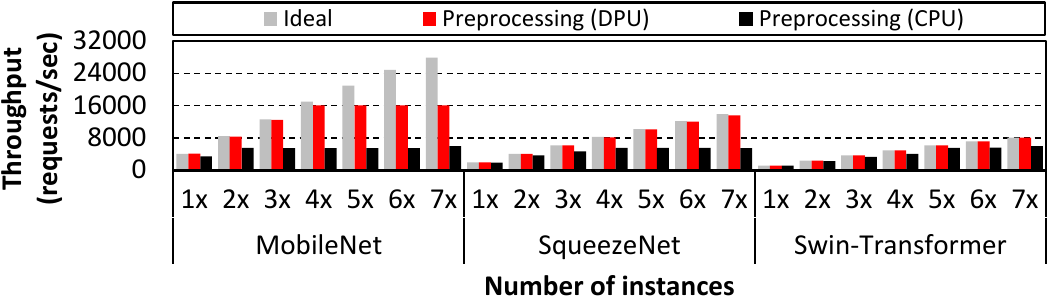}}\\
\vspace{-0.4em}
\subfloat[Audio processing]{\includegraphics[width=0.47\textwidth]{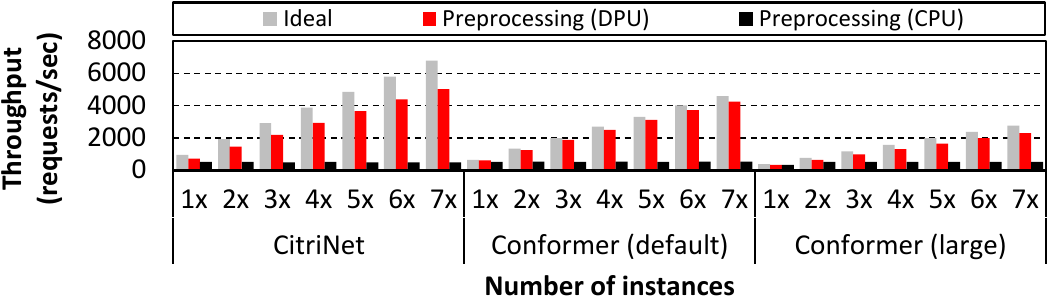}}
\vspace{-0.4em}
\caption{Inference throughput of \migsmall when the number of \emph{activated} inference servers is changed from one ($1\times$) to seven ($7\times$).}
\label{fig:eval_throughput}
\vspace{-0.8em}
\end{figure}

{\bf Throughput.} \fig{fig:eval_throughput} shows the end-to-end inference throughput of our studied systems. As depicted, the performance of the baseline system quickly gets bottlenecked by the data preprocessing stage and suffers from an average of $77.2\%$ of throughput loss vs. ``Ideal''. Our \proposed, on the other hand, achieves more than $91.6\%$ of the performance of ``Ideal'' for $5$ out of the $6$ studied models, providing an average $3.7\times$ end-to-end throughput improvement vs. baseline. 

\begin{figure}[t!] \centering
\subfloat[Computer vision]{\includegraphics[width=0.48\textwidth]{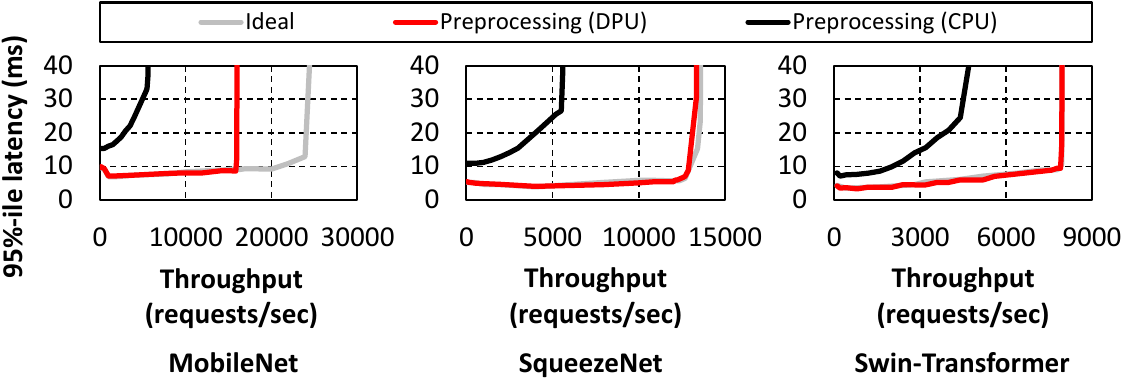}}\\
\vspace{-0.4em}
\subfloat[Audio processing]{\includegraphics[width=0.48\textwidth]{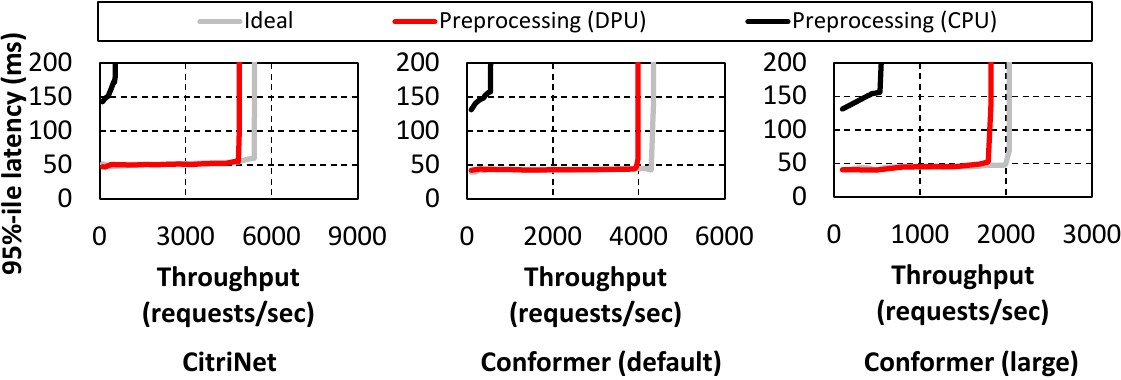}}
\vspace{-0.4em}
\caption{Throughput vs. tail latency curve. All three designs run on top of a \migsmall MIG configuration.}
\label{fig:eval_throughput_latency}
\end{figure}

\begin{figure}[t!] \centering
\includegraphics[width=0.47\textwidth]{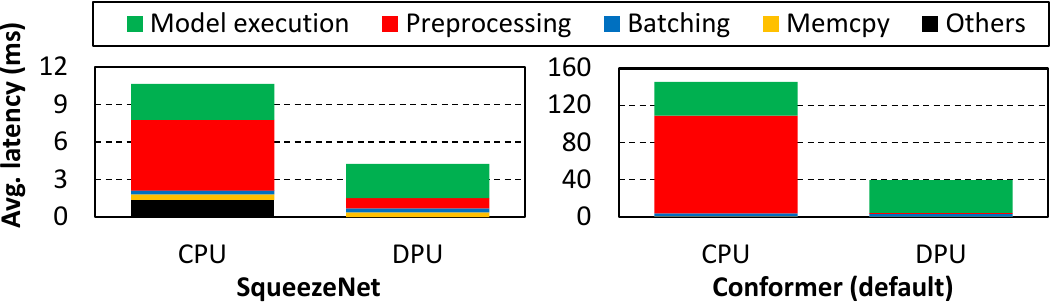}
\vspace{-0.4em}
\caption{Breakdown of end-to-end latency while running the experiments shown in \fig{fig:eval_throughput_latency}. For brevity, we only show results for SqueezeNet (left) and Conformer(default) (right). }
\label{fig:eval_latency_breakdown}
\end{figure}

{\bf Latency.} We also show how well \proposed can sustain a much higher throughput while staying within a tail latency target (\fig{fig:eval_throughput_latency}).  In general, the baseline system	experiences a sharp increase in tail latency at a much lower throughput than the other two designs.  This is because of the large
latency overhead incurred during data preprocessing (e.g., $53\%$ and $72\%$ of inference time in SqueezeNet and Conformer(default), see	\fig{fig:eval_latency_breakdown}), suffering from high tail latency even when the inference server is handling a small number of incoming requests.  Our \proposed system completely resolves this performance bottleneck and shows that it can sustain $91.6\%$ of the throughput of ``Ideal'' while staying within similar tail latency boundaries. This is also demonstrated by how close the red-colored lines (\proposed) are to the gray-colored lines (``Ideal'') for $5$ out of the $6$ studied workloads.  

\begin{figure}[t!] \centering
\vspace{-0.4em}
\subfloat[Computer vision]{\includegraphics[width=0.47\textwidth]{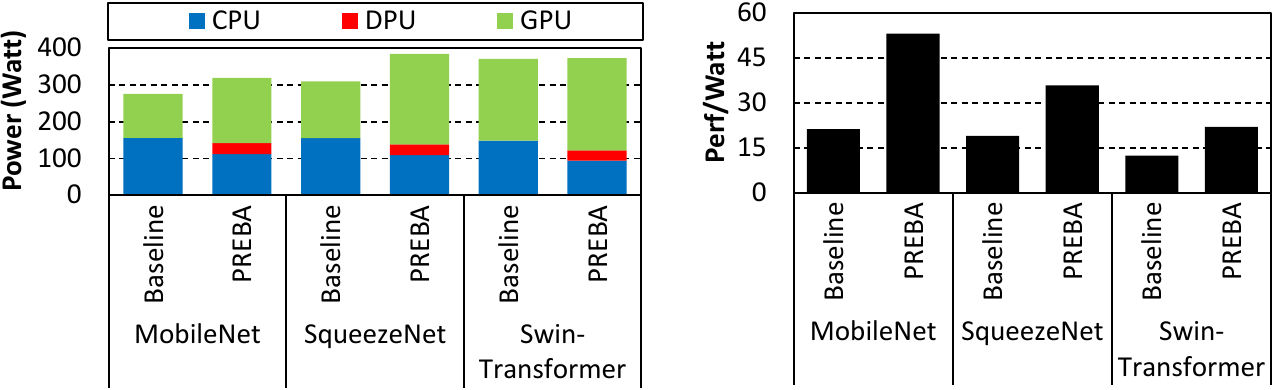}}\\
\vspace{-0.4em}
\subfloat[Audio processing]{\includegraphics[width=0.47\textwidth]{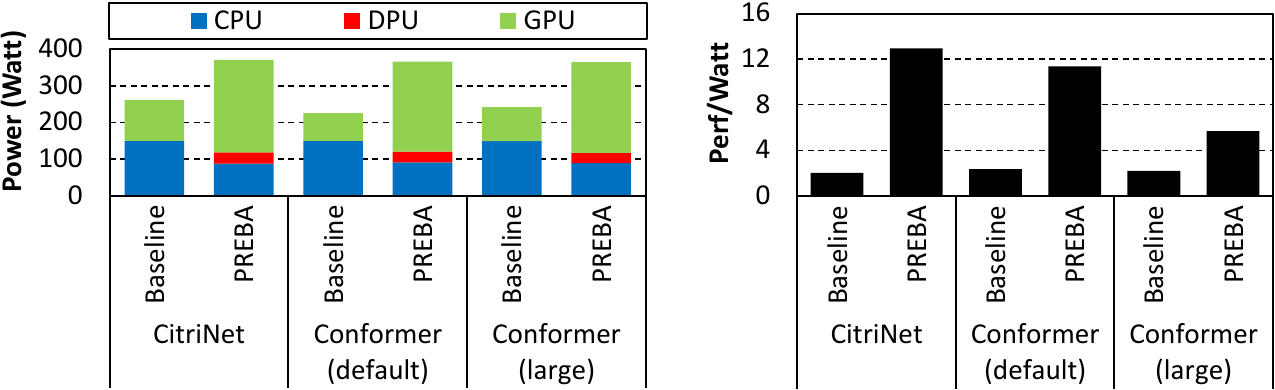}}
\vspace{-0.4em}
\caption{Power consumption (left) and energy-efficiency (right).}
\label{fig:eval_power}
\vspace{-1.2em}
\end{figure}

\begin{figure}[t!] \centering
\subfloat[Computer vision]{\includegraphics[width=0.23\textwidth]{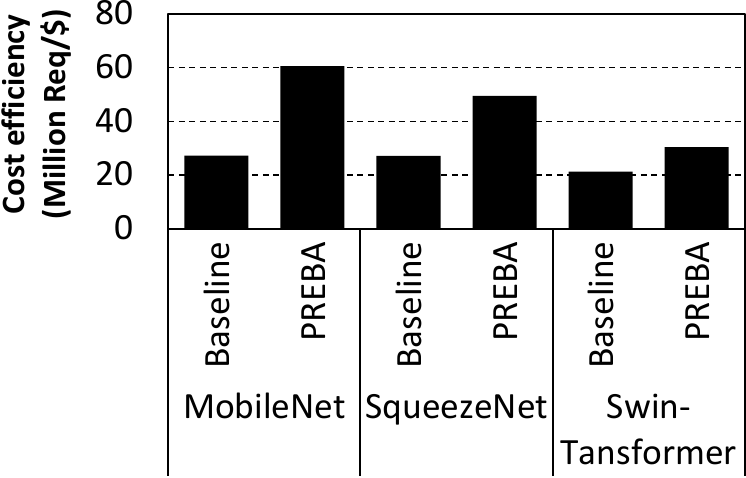}}
\hspace{1 mm}
\subfloat[Audio processing]{\includegraphics[width=0.23\textwidth]{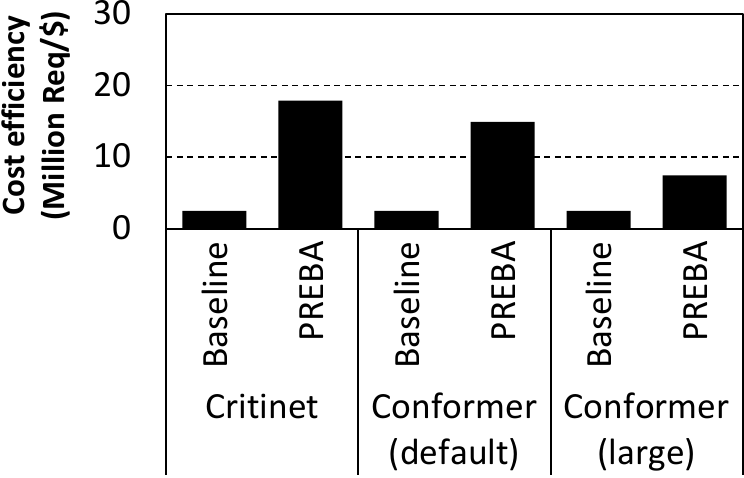}}\\
\vspace{-0.8em}
\caption{PREBA's cost-efficiency (TCO). }
\label{fig:eval_TCO}
\end{figure}

\subsection{Power and Energy-Efficiency}
\label{sect:eval_energy}
\label{page:TCO_2}
We now evaluate \proposed's effect on power consumption and energy-efficiency (Perf/Watt). In \fig{fig:eval_power}, we provide a breakdown of baseline vs. \proposed's system-wide power consumption (left) and how that translates into energy-efficiency (right). In general, \proposed's DPU incurs an additional power consumption but makes up for that overhead by reducing power consumed by the CPU (an average $35.4\%$ reduction in CPU power consumption). What is noteworthy is that \proposed's  GPU power consumption has increased by an average of $2.8\times$ for the three audio processing models. This is an artifact of the significant increase in  GPU utilization enabled by our DPU architecture as it successfully unleashes the CPU-side data preprocessing bottlenecks of the baseline system. Because \proposed provides significant end-to-end speedup (i.e., reduction in execution time), the system-wide energy-efficiency is improved by an average of $3.5\times$.

\subsection{Cost-Efficiency (TCO)}
\label{sect:eval_tco}

To evaluate PREBA's effect on reducing TCO, we quantify cost-efficiency using the evaluation metric suggested by prior work: (\mbox{$\frac{\textit{Throughput} \times \textit{time}}{\text{CAPEX} + \text{OPEX}}$~\cite{e3_tco}}). CAPEX (CAPital EXpenditure) denotes the one-time cost to purchase all hardware components, including server node~\cite{supermicro}, GPU and~\cite{a100cost}, FPGA~\cite{FPGAcost}. Also, \textit{time} is the active duration of these hardware components (3 years). Finally, OPEX (OPerating EXpence) encompasses the total power expense for operation, calculated as the product of power consumption, \textit{time}, and the cost of electricity (\mbox{$\$0.139/khW$}\mbox{\cite{electriccost}}). Our proposed system outperforms the baseline in all evaluated workloads and demonstrates an average of $3.0\times$ better cost efficiency, as shown in \fig{fig:eval_TCO}.

\subsection{Ablation Study}
\label{sect:eval_ablation}
\label{page:breakdown}

\begin{figure}[t!] \centering
\vspace{+0.4em}
\includegraphics[width=0.47\textwidth]{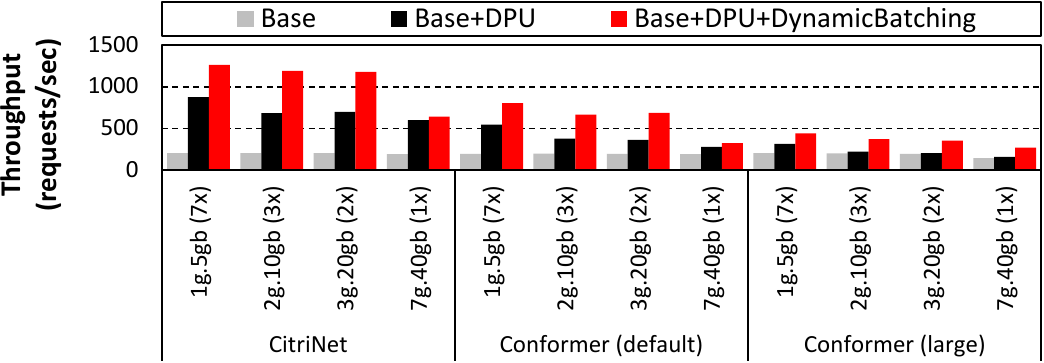}\\
\caption{Ablation study showing the effectiveness of PREBA's hardware and software proposals. Because the dynamic batching system targets audio processing algorithms, we only present speedup numbers for our audio processing AI workloads.}
\label{fig:eval_sensitivity_batching}
\end{figure}

The PREBA design point presented so far assumed that ``all'' of our proposal (i.e., DPU and dynamic batching system)
are applied on top of the baseline system.
In \fig{fig:eval_sensitivity_batching}, we conduct an ablation study to quantitatively
evaluate the effectiveness of each of our hardware (DPU) and software (dynamic batching system)
proposal individually. The leftmost gray bar (``Base'') represents the default baseline system without any of PREBA's DPU and dynamic batching system incorporated.
The ``Base+DPU'' design point (black bar) is one where ``Base'' is enhanced with our DPU to accelerate data preprocessing but without
dynamic batching. Finally, ``Base+DPU+DynamicBatching'' (red bar) refers to our complete PREBA system incorporating both
of our hardware/software proposals.
Overall, ``Base+DPU'' outperforms ``Base'' by an average $101\%$.
Adding our dynamic batching system to ``Base+DPU'' provides further performance improvement by an average $54\%$, demonstrating
the efficacy and importance of each of our proposal.

\section{Conclusion}
\label{sect:conclusion}

This paper presents a hardware/software co-design named \proposed which
is a high-performance AI inference server targeting NVIDIA's MIG architecture.
Our characterization uncovers the significant data
preprocessing bottlenecks MIG incurs and how MIG affects the inference
server's batching system. Motivated by such, we propose a
DPU architecture that addresses the preprocessing bottlenecks of
MIG inference. We then develop a dynamic batching system tailored to
the unique algorithmic properties of MIG inference servers as well as the AI
models running on top of it.  Using commodity hardware and open-source
software, we demonstrated the merits of \proposed over state-of-the-art solutions.

\bibliographystyle{plain}
\bibliography{refs}

\begin{thebibliography}{10}

\bibitem{graphcore}
{GRAPHCORE}.
\newblock \url{https://www.graphcore.ai/products/c600}.

\bibitem{librosa}
{LibrosaCpp}.
\newblock \url{https://github.com/ewan-xu/LibrosaCpp}.

\bibitem{opencv}
{OpenCV}.
\newblock \url{https://opencv.org/}.

\bibitem{rebellions}
{rebellions}.
\newblock \url{https://rebellions.ai/rebellions-product/atom-2/}.

\bibitem{dcs}
J.~Ahn, D.~Kwon, Y.~Kim, M.~Ajdari, J.~Lee, and J.~Kim.
\newblock {DCS: A Fast and Scalable Device-Centric Server Architecture}.
\newblock In {\em Proceedings of the International Symposium on
  Microarchitecture (MICRO)}, December 2015.

\bibitem{cnvlutin}
J.~Albericio, P.~Judd, T.~Hetherington, T.~Aamodt, N.~E. Jerger, and
  A.~Moshovos.
\newblock {Cnvlutin: Ineffectual-Neuron-Free Deep Convolutional Neural Network
  Computing}.
\newblock In {\em Proceedings of the International Symposium on Computer
  Architecture (ISCA)}, June 2016.

\bibitem{a100cost}
AMAZON.
\newblock
  \url{https://www.amazon.com/NVIDIA-Tesla-A100-Ampere-Graphics/dp/B0BGZJ27SL},
  2024.

\bibitem{u55c}
{AMD Xilinx}.
\newblock {Alveo U55C High Performance Compute Card}.
\newblock
  \url{https://www.xilinx.com/products/boards-and-kits/alveo/u55c.html}, 2023.

\bibitem{vitis}
{AMD Xilinx}.
\newblock {AMD Vitis}.
\newblock
  \url{https://www.xilinx.com/products/design-tools/vitis/vitis-hls.html},
  2023.

\bibitem{vitis_libraries}
{AMD Xilinx}.
\newblock {Vitis Libraries}.
\newblock \url{https://docs.amd.com/r/en-US/Vitis_Libraries}, 2024.

\bibitem{ankit2019puma}
Aayush Ankit, Izzat~El Hajj, Sai~Rahul Chalamalasetti, Geoffrey Ndu, Martin
  Foltin, R.~Stanley Williams, Paolo Faraboschi, Wen mei Hwu, John~Paul
  Strachan, Kaushik Roy, and Dejan~S Milojicic.
\newblock {PUMA: A Programmable Ultra-efficient Memristor-based Accelerator for
  Machine Learning Inference}.
\newblock 2019.

\bibitem{arashloo2020enabling}
Mina~Tahmasbi Arashloo, Alexey Lavrov, Manya Ghobadi, Jennifer Rexford, David
  Walker, and David Wentzlaff.
\newblock {Enabling Programmable Transport Protocols in High-Speed NICs}.
\newblock In {\em USENIX Symposium on Networked Systems Design and
  Implementation (NSDI)}, 2020.

\bibitem{amazon_nitro}
{AWS}.
\newblock {AWS Nitro System}.
\newblock \url{https://aws.amazon.com/ec2/nitro/}, 2023.

\bibitem{inferentia}
A.W.Services.
\newblock {AWS inferentia}.
\newblock \url{https://aws.amazon.com/machine-learning/inferentia/}, 2024.

\bibitem{fungible}
Girish Bablani.
\newblock {Microsoft Announces Acquisition of Fungible to Accelerate Datacenter
  Innovation}.
\newblock \url{https://www.fungible.com}, 2023.

\bibitem{boo:2023:f4t}
Junehyuk Boo, Yujin Chung, Eunjin Baek, Seongmin Na, Changsu Kim, and Jangwoo
  Kim.
\newblock {F4T: A Fast and Flexible FPGA-based Full-stack TCP Acceleration
  Framework}.
\newblock In {\em Proceedings of the International Symposium on Computer
  Architecture (ISCA)}, 2023.

\bibitem{rodinia}
Shuai Che, Michael Boyer, Jiayuan Meng, David Tarjan, Jeremy~W. Sheaffer,
  Sang-Ha Lee, and Kevin Skadron.
\newblock {Rodinia: A Benchmark Suite for Heterogeneous Computing}.
\newblock In {\em Proceedings of the International Symposium on Workload
  Characterization (IISWC)}, 2009.

\bibitem{Prophet:2017}
Quan Chen, Hailong Yang, Minyi Guo, Ram~Srivatsa Kannan, Jason Mars, and
  Lingjia Tang.
\newblock {Prophet: Precise QoS Prediction on Non-Preemptive Accelerators to
  Improve Utilization in Warehouse-Scale Computers}.
\newblock In {\em Proceedings of the International Conference on Architectural
  Support for Programming Languages and Operation Systems (ASPLOS)}, 2017.

\bibitem{Baymax:2016}
Quan Chen, Hailong Yang, Jason Mars, and Lingjia Tang.
\newblock {Baymax: QoS Awareness and Increased Utilization for Non-Preemptive
  Accelerators in Warehouse Scale Computers}.
\newblock In {\em Proceedings of the International Conference on Architectural
  Support for Programming Languages and Operation Systems (ASPLOS)}, 2016.

\bibitem{eyeriss}
Y.~Chen, T.~Krishna, J.~Emer, and V.~Sze.
\newblock {Eyeriss: An Energy-Efficient Reconfigurable Accelerator for Deep
  Convolutional Neural Networks}.
\newblock In {\em Proceedings of the International Solid State Circuits
  Conference (ISSCC)}, February 2016.

\bibitem{chen2023bm}
Yiquan Chen, Jiexiong Xu, Chengkun Wei, Yijing Wang, Xin Yuan, Yangming Zhang,
  Xulin Yu, Yi~Chen, Zeke Wang, Shuibing He, and Wenzhi Chen.
\newblock {BM-Store: A Transparent and High-performance Local Storage
  Architecture for Bare-metal Clouds Enabling Large-scale Deployment}.
\newblock In {\em Proceedings of the International Symposium on
  High-Performance Computer Architecture (HPCA)}, 2023.

\bibitem{cheng2019dlbooster}
Yang Cheng, Dan Li, Zhiyuan Guo, Binyao Jiang, Jiaxin Lin, Xi~Fan, Jinkun Geng,
  Xinyi Yu, Wei Bai, Lei Qu, Ran Shu, Peng Cheng, Yongqiang Xiong, and Jianping
  Wu.
\newblock {Dlbooster: Boosting End-to-End Deep Learning Workflows with
  Offloading Data Preprocessing Pipelines}.
\newblock In {\em Proceedings International Conference on Parallel Processing},
  2019.

\bibitem{lazybatching}
Yujeong Choi, Yunseong Kim, and Minsoo Rhu.
\newblock {Lazy Batching: An SLA-aware Batching System for Cloud Machine
  Learning Inference}.
\newblock In {\em Proceedings of the International Symposium on
  High-Performance Computer Architecture (HPCA)}, 2021.

\bibitem{prema}
Yujeong Choi and Minsoo Rhu.
\newblock {PREMA: A Predictive Multi-task Scheduling Algorithm For Preemptible
  Neural Processing Units}.
\newblock In {\em Proceedings of the International Symposium on
  High-Performance Computer Architecture (HPCA)}, 2020.

\bibitem{clipper}
Daniel Crankshaw, Xin Wang, Guilio Zhou, Michael~J Franklin, Joseph~E Gonzalez,
  and Ion Stoica.
\newblock {Clipper: A Low-Latency Online Prediction Serving System}.
\newblock In {\em Proceedings of USENIX Symposium on Networked Systems Design
  and Implementation (NSDI)}, 2017.

\bibitem{electriccost}
DATACENTERS.com.
\newblock Everything you need to know about data center power.
\newblock
  \url{https://www.datacenters.com/news/everything-you-need-to-know-about-data-center-power},
  2020.

\bibitem{Neuralcache:isca:2018}
Charles Eckert, Xiaowei Wang, Jingcheng Wang, Arun Subramaniyan, Ravi Iyer,
  Dennis Sylvester, David Blaauw, and Reetuparna Das.
\newblock {Neural Cache: Bit-serial In-cache Acceleration of Deep Neural
  Networks}.
\newblock In {\em Proceedings of the International Symposium on Computer
  Architecture (ISCA)}, 2018.

\bibitem{MAICC:micro:2023}
Renhao Fan, Yikai Cui, Qilin Chen, Mingyu Wang, Youhui Zhang, Weimin Zheng, and
  Zhaolin Li.
\newblock {MAICC : A Lightweight Many-core Architecture with In-Cache Computing
  for Multi-DNN Parallel Inference}.
\newblock In {\em Proceedings of the International Symposium on
  Microarchitecture (MICRO)}, 2023.

\bibitem{cellular_batching}
Pin Gao, Lingfan Yu, Yongwei Wu, and Jinyang Li.
\newblock {Low Latency RNN Inference with Cellular Batching}.
\newblock In {\em Proceedings of the EuroSys Conference}, 2018.

\bibitem{Batchmaker}
Pin Gao, Lingfan Yu, Yongwei Wu, and Jinyang Li.
\newblock {Low Latency RNN Inference with Cellular Batching}.
\newblock In {\em Proceedings of the EuroSys Conference (EuroSys)}, 2018.

\bibitem{graur2022cachew}
Dan Graur, Damien Aymon, Dan Kluser, Tanguy Albrici, Chandramohan~A Thekkath,
  and Ana Klimovic.
\newblock {Cachew: Machine Learning Input Data Processing as a Service}.
\newblock In {\em USENIX Annual Technical Conference (USENIX ATC)}, 2022.

\bibitem{conformer}
Anmol Gulati, James Qin, Chung-Cheng Chiu, Niki Parmar, Yu~Zhang, Jiahui Yu,
  Wei Han, Shibo Wang, Zhengdong Zhang, and Yonghui Wu.
\newblock {Conformer: Convolution-augmented Transformer for Speech
  Recognition}.
\newblock 2020.

\bibitem{song:2015:eie}
S.~Han, X.~Liu, H.~Mao, J.~Pu, A.~Pedram, M.~Horowitz, and W.~Dally.
\newblock {EIE: Efficient Inference Engine on Compressed Deep Neural Network}.
\newblock In {\em Proceedings of the International Symposium on Computer
  Architecture (ISCA)}, June 2016.

\bibitem{djinn_and_tonic}
J.~Hauswald, Y.~Kang, M.~A. Laurenzano, Q.~Chen, C.~Li, T.~Mudge, R.~G.
  Dreslinski, J.~Mars, and L.~Tang.
\newblock {DjiNN and Tonic: DNN as a Service and Its Implications for Future
  Warehouse Scale Computers}.
\newblock In {\em Proceedings of the International Symposium on Computer
  Architecture (ISCA)}, June 2015.

\bibitem{mobilenet}
Andrew Howard, Mark Sandler, Grace Chu, Liang-Chieh Chen, Bo~Chen, Mingxing
  Tan, Weijun Wang, Yukun Zhu, Ruoming Pang, Vijay Vasudevan, Quoc~V. Le, and
  Hartwig Adam.
\newblock {Searching for MobileNetV3}.
\newblock {\em arXiv preprint arXiv:1905.02244}, 2019.

\bibitem{hbm3e}
Hynix.
\newblock {SK hynix Develops World’s Best Performing HBM3E, Provides Samples
  to Customer for Performance Evaluation}.
\newblock 2023.

\bibitem{squeezenet}
Forrest~N Iandola, Song Han, Matthew~W Moskewicz, Khalid Ashraf, William~J
  Dally, and Kurt Keutzer.
\newblock {SqueezeNet: AlexNet-level Accuracy with 50x Fewer Parameters and
  $<$0.5MB Model Size}.
\newblock {\em arXiv preprint arXiv:1602.07360}, 2016.

\bibitem{ilsvrc2012}
{ImageNet}.
\newblock {ImageNet Large Scale Visual Recognition Challenge 2012
  (ILSVRC2012)}.
\newblock \url{https://www.image-net.org/challenges/LSVRC/2012/}, 2012.

\bibitem{intel_ipu}
{Intel}.
\newblock {Intel Infrastructure Processing Unit (Intel IPU)}.
\newblock
  \url{https://www.intel.com/content/www/us/en/products/details/network-io/ipu.html},
  2023.

\bibitem{tpu_paper}
N.~P. Jouppi, C.~Young, N.~Patil, D.~Patterson, G.~Agrawal, R.~Bajwa, S.~Bates,
  S.~Bhatia, N.~Boden, A.~Borchers, R.~Boyle, P.~Cantin, C.~Chao, C.~Clark,
  J.~Coriell, M.~Daley, M.~Dau, J.~Dean, B.~Gelb, T.~V. Ghaemmaghami,
  R.~Gottipati, W.~Gulland, R.~Hagmann, C.~R. Ho, D.~Hogberg, J.~Hu, R.~Hundt,
  D.~Hurt, J.~Ibarz, A.~Jaffey, A.~Jaworski, A.~Kaplan, H.~Khaitan,
  D.~Killebrew, A.~Koch, N.~Kumar, S.~Lacy, J.~Laudon, J.~Law, D.~Le, C.~Leary,
  Z.~Liu, K.~Lucke, A.~Lundin, G.~MacKean, A.~Maggiore, M.~Mahony, K.~Miller,
  R.~Nagarajan, R.~Narayanaswami, R.~Ni, K.~Nix, T.~Norrie, M.~Omernick,
  N.~Penukonda, A.~Phelps, J.~Ross, M.~Ross, A.~Salek, E.~Samadiani, C.~Severn,
  G.~Sizikov, M.~Snelham, J.~Souter, D.~Steinberg, A.~Swing, M.~Tan,
  G.~Thorson, B.~Tian, H.~Toma, E.~Tuttle, V.~Vasudevan, R.~Walter, W.~Wang,
  E.~Wilcox, and D.~H. Yoon.
\newblock {In-datacenter Performance Analysis of A Tensor Processing Unit}.
\newblock In {\em Proceedings of the International Symposium on Computer
  Architecture (ISCA)}, 2017.

\bibitem{opencl}
{Khronos Group}.
\newblock {OpenCL}.
\newblock \url{https://www.khronos.org/opencl/}, 2023.

\bibitem{kim2023rearchitecting}
Taehyun Kim, Deondre~Martin Ng, Junzhi Gong, Youngjin Kwon, Minlan Yu, and
  KyoungSoo Park.
\newblock {Rearchitecting the TCP Stack for I/O-Offloaded Content Delivery}.
\newblock In {\em USENIX Symposium on Networked Systems Design and
  Implementation (NSDI)}, 2023.

\bibitem{paris_and_elsa}
Yunseong Kim, Yujeong Choi, and Minsoo Rhu.
\newblock {PARIS and ELSA: An Elastic Scheduling Algorithm for Reconfigurable
  Multi-GPU Inference Servers}.
\newblock In {\em Proceedings of the ACM/IEEE Design Automation Conference
  (DAC)}, 2022.

\bibitem{dcs_ctrl}
Dongup Kwon, Jaehyung Ahn, Dongju Chae, Mohammadamin Ajdari, Jaewon Lee, Suheon
  Bae, Youngsok Kim, and Jangwoo Kim.
\newblock {Dcs-ctrl: A Fast and Flexible Device-Control Mechanism for
  Device-Centric Server Architecture}.
\newblock In {\em Proceedings of the International Symposium on Computer
  Architecture (ISCA)}, 2018.

\bibitem{kwon:2020:osdi}
Dongup Kwon, Junehyuk Boo, Dongryeong Kim, and Jangwoo Kim.
\newblock {FVM: FPGA-assisted Virtual Device Emulation for Fast, Scalable, and
  Flexible Storage Virtualization}.
\newblock In {\em USENIX Symposium on Operating Systems Design and
  Implementation (OSDI)}, 2020.

\bibitem{kwon:2021:atc}
Dongup Kwon, Dongryeong Kim, Junehyuk Boo, Wonsik Lee, and Jangwoo Kim.
\newblock {A Fast and Flexible Hardware-based Virtualization Mechanism for
  Computational Storage Devices}.
\newblock In {\em USENIX Annual Technical Conference (USENIX ATC)}, 2021.

\bibitem{li2022characterizing}
Baolin Li, Viiay Gadepally, Siddharth Samsi, and Devesh Tiwari.
\newblock {Characterizing Multi-Instance GPU for Machine Learning Workloads}.
\newblock In {\em Proceedings of IEEE International Parallel and Distributed
  Processing Symposium Workshops (IPDPSW)}, 2022.

\bibitem{miso}
Baolin Li, Tirthak Patel, Siddharth Samsi, Vijay Gadepally, and Devesh Tiwari.
\newblock {MISO: Exploiting Multi-Instance GPU Capability on Multi-Tenant GPU
  Clusters}.
\newblock In {\em Proceedings of the Symposium on Cloud Computing (SoCC)},
  2022.

\bibitem{li2020leapio}
Huaicheng Li, Mingzhe Hao, Stanko Novakovic, Vaibhav Gogte, Sriram Govindan,
  Dan~RK Ports, Irene Zhang, Ricardo Bianchini, Haryadi~S Gunawi, and Anirudh
  Badam.
\newblock {Leapio: Efficient and Portable Virtual NVMe Storage on ARM SOCs}.
\newblock In {\em Proceedings of the International Conference on Architectural
  Support for Programming Languages and Operation Systems (ASPLOS)}, 2020.

\bibitem{e3_tco}
Ming Liu, Simon Peter, Arvind Krishnamurthy, and Phitchaya~Mangpo
  Phothilimthana.
\newblock {E3: Energy-efficient Microservices on SmartNIC-accelerated Servers}.
\newblock In {\em USENIX Annual Technical Conference (USENIX ATC)}, 2019.

\bibitem{swin}
Ze~Liu, Yutong Lin, Yue Cao, Han Hu, Yixuan Wei, Zheng Zhang, Stephen Lin, and
  Baining Guo.
\newblock {Swin Transformer: Hierarchical Vision Transformer Using Shifted
  Windows}.
\newblock In {\em Proceedings of the International Conference on Computer
  Vision (ICCV)}, 2021.

\bibitem{citrinet}
Somshubra Majumdar, Jagadeesh Balam, Oleksii Hrinchuk, Vitaly Lavrukhin, Vahid
  Noroozi, and Boris Ginsburg.
\newblock {Citrinet: Closing the Gap between Non-Autoregressive and
  Autoregressive End-to-End Models for Automatic Speech Recognition}, 2021.

\bibitem{mangoboost}
{MangoBoost}.
\newblock {Boost Your Datacenter!}
\newblock \url{https://mangoboost.io/}, 2023.

\bibitem{mita}
Meta.
\newblock {Meta MITA v1}.
\newblock
  \url{https://ai.meta.com/blog/meta-training-inference-accelerator-AI-MTIA/},
  2023.

\bibitem{min2021gimbal}
Jaehong Min, Ming Liu, Tapan Chugh, Chenxingyu Zhao, Andrew Wei, In~Hwan Doh,
  and Arvind Krishnamurthy.
\newblock {Gimbal: Enabling Multi-Tenant Storage Disaggregation on SmartNIC
  JBOFs}.
\newblock In {\em Proceedings of the ACM SIGCOMM Conference}, 2021.

\bibitem{mohan2021analyzing}
Jayashree Mohan, Amar Phanishayee, Ashish Raniwala, and Vijay Chidambaram.
\newblock {Analyzing and Mitigating Data Stalls in DNN Training}.
\newblock {\em Proceedings of the VLDB Endowment}, 2021.

\bibitem{moon2020acceltcp}
YoungGyoun Moon, SeungEon Lee, Muhammad~Asim Jamshed, and KyoungSoo Park.
\newblock {AccelTCP: Accelerating Network Applications with Stateful TCP
  Offloading}.
\newblock In {\em USENIX Symposium on Networked Systems Design and
  Implementation (NSDI)}, 2020.

\bibitem{trtis}
{NVIDIA}.
\newblock {NVIDIA Triton Inference Server}.
\newblock \url{https://developer.nvidia.com/nvidia-triton-inference-server}.

\bibitem{tensorrt}
NVIDIA.
\newblock {NVIDIA TensorRT: Programmable Inference Accelerator}.
\newblock 2018.

\bibitem{t4}
NVIDIA.
\newblock {NVIDIA T4}.
\newblock
  \url{https://www.nvidia.com/content/dam/en-zz/Solutions/Data-Center/tesla-t4/t4-tensor-core-datasheet-951643.pdf},
  2019.

\bibitem{a100}
NVIDIA.
\newblock {NVIDIA A100}.
\newblock
  \url{https://www.nvidia.com/content/dam/en-zz/Solutions/Data-Center/a100/pdf/nvidia-a100-datasheet.pdf},
  2020.

\bibitem{cuda}
{NVIDIA}.
\newblock {NVIDIA CUDA Programming Guide}, 2021.

\bibitem{mig}
{NVIDIA}.
\newblock {Multi-Instance GPU}.
\newblock \url{https://www.nvidia.com/en-us/technologies/multi-instance-gpu/},
  2023.

\bibitem{bluefield}
{NVIDIA}.
\newblock {NVIDIA BlueField Data Processing Units}.
\newblock
  \url{https://www.nvidia.com/en-us/networking/products/data-processing-unit/},
  2023.

\bibitem{nvidia_nemo}
{NVIDIA}.
\newblock {NVIDIA NeMo}.
\newblock
  \url{https://www.nvidia.com/en-us/ai-data-science/generative-ai/nemo-framework/},
  2023.

\bibitem{tf_serving}
Christopher Olston, Noah Fiedel, Kiril Gorovoy, Jeremiah Harmsen, Li~Lao,
  Fangwei Li, Vinu Rajashekhar, Sukriti Ramesh, and Jordan Soyke.
\newblock {TensorFlow-Serving: Flexible, High-Performance ML Serving}.
\newblock {\em arXiv preprint arXiv:1712.06139}, 2017.

\bibitem{librispeech}
{Open SLR}.
\newblock {LibriSpeech ASR corpus}.
\newblock \url{http://www.openslr.org/12/}.

\bibitem{scnn}
A.~Parashar, M.~Rhu, A.~Mukkara, A.~Puglielli, R.~Venkatesan, B.~Khailany,
  J.~Emer, S.~W. Keckler, and W.~J. Dally.
\newblock {SCNN: An Accelerator for Compressed-sparse Convolutional Neural
  Networks}.
\newblock In {\em Proceedings of the International Symposium on Computer
  Architecture (ISCA)}, June 2017.

\bibitem{park2020trainbox}
Pyeongsu Park, Heetaek Jeong, and Jangwoo Kim.
\newblock {TrainBox: An Extreme-Scale Neural Network Training Server
  Architecture by Systematically Balancing Operations}.
\newblock In {\em Proceedings of the International Symposium on
  Microarchitecture (MICRO)}, 2020.

\bibitem{sr_iov}
{PCI SIG}.
\newblock {Single-Root Input/Output Virtualization}.
\newblock \url{https://www.pcisig.com/specifications}.

\bibitem{pytorch_audiopreproc}
{PyTorch}.
\newblock {Torchaudio Transforms}.
\newblock \url{https://pytorch.org/audio/stable/transforms.html}.

\bibitem{pytorch_imagepreproc}
{PyTorch}.
\newblock {Transforming and Augmenting Images}.
\newblock \url{https://pytorch.org/vision/stable/transforms.html}.

\bibitem{torch_hub}
{PyTorch}.
\newblock {PyTorch Hub}.
\newblock \url{https://pytorch.org/hub/}, 2023.

\bibitem{mlperf_inf}
Vijay~Janapa Reddi, Christine Cheng, David Kanter, Peter Mattson, Guenther
  Schmuelling, Carole-Jean Wu, Brian Anderson, Maximilien Breughe, Mark
  Charlebois, William Chou, Ramesh Chukka, Cody Coleman, Sam Davis, Pan Deng,
  Greg Diamos, Jared Duke, Dave Fick, J.~Scott Gardner, Itay Hubara, Sachin
  Idgunji, Thomas~B. Jablin, Jeff Jiao, Tom~St. John, Pankaj Kanwar, David Lee,
  Jeffery Liao, Anton Lokhmotov, Francisco Massa, Peng Meng, Paulius
  Micikevicius, Colin Osborne, Gennady Pekhimenko, Arun Tejusve~Raghunath
  Rajan, Dilip Sequeira, Ashish Sirasao, Fei Sun, Hanlin Tang, Michael Thomson,
  Frank Wei, Ephrem Wu, Lingjie Xu, Koichi Yamada, Bing Yu, George Yuan, Aaron
  Zhong, Peizhao Zhang, and Yuchen Zhou.
\newblock {MLPerf Inference Benchmark}.
\newblock In {\em Proceedings of the International Symposium on Computer
  Architecture (ISCA)}, 2020.

\bibitem{robroek2022analysis}
Ties Robroek, Ehsan Yousefzadeh-Asl-Miandoab, and P{\i}nar T{\"o}z{\"u}n.
\newblock {An Analysis of Collocation on GPUs for Deep Learning Training}.
\newblock {\em arXiv preprint arXiv:2209.06018}, 2022.

\bibitem{gddr7}
Samsung.
\newblock {Samsung Develops Industry’s First GDDR7 DRAM To Unlock the Next
  Generation of Graphics Performance}.
\newblock 2023.

\bibitem{Simba:micro:2019}
Yakun~Sophia Shao, Jason Clemons, Rangharajan Venkatesan, Brian Zimmer, Matthew
  Fojtik, Nan Jiang, Ben Keller, Alicia Klinefelter, Nathaniel Pinckney,
  Priyanka Raina, Stephen~G. Tell, Yanqing Zhang, William~J. Dally, Joel Emer,
  C.~Thomas Gray, Brucek Khailany, and Stephen~W. Keckler.
\newblock {Simba: Scaling Deep-Learning Inference with Multi-Chip-Module-Based
  Architecture}.
\newblock In {\em Proceedings of the International Symposium on
  Microarchitecture (MICRO)}, 2019.

\bibitem{laconic:isca:2019}
Sayeh Sharify, Alberto~Delmas Lascorz, Mostafa Mahmoud, Milos Nikolic, Kevin
  Siu, Dylan~Malone Stuart, Zissis Poulos, and Andreas Moshovos.
\newblock {Laconic Deep Learning Inference Acceleration}.
\newblock In {\em Proceedings of the International Symposium on Computer
  Architecture (ISCA)}, 2019.

\bibitem{bitfusion:isca:2018}
Hardik Sharma, Jongse Park, Naveen Suda, Liangzhen Lai, Benson Chau, Joon~Kyung
  Kim, Vikas Chandra, and Hadi Esmaeilzadeh.
\newblock {Bit Fusion: Bit-Level Dynamically Composable Architecture for
  Accelerating Deep Neural Network}.
\newblock In {\em Proceedings of the International Symposium on Computer
  Architecture (ISCA)}, 2018.

\bibitem{shashidhara2022flextoe}
Rajath Shashidhara, Tim Stamler, Antoine Kaufmann, and Simon Peter.
\newblock {FlexTOE: Flexible TCP Offload with Fine-Grained Parallelism}.
\newblock In {\em USENIX Symposium on Networked Systems Design and
  Implementation (NSDI)}, 2022.

\bibitem{Maia}
Jake Siegel.
\newblock {Microsoft Maia v1}.
\newblock
  \url{https://news.microsoft.com/source/features/ai/in-house-chips-silicon-to-service-to-meet-ai-demand/},
  2023.

\bibitem{supermicro}
SuperMICRO.
\newblock
  \url{https://store.supermicro.com/us_en/mainstream-amd-2u-as-2024s-tr.html},
  2024.

\bibitem{tork2020lynx}
Maroun Tork, Lina Maudlej, and Mark Silberstein.
\newblock {Lynx: A SmartNIC-Driven Accelerator-Centric Architecture for Network
  Servers}.
\newblock In {\em Proceedings of the International Conference on Architectural
  Support for Programming Languages and Operation Systems (ASPLOS)}, 2020.

\bibitem{um2023fastflow}
Taegeon Um, Byungsoo Oh, Byeongchan Seo, Minhyeok Kweun, Goeun Kim, and
  Woo-Yeon Lee.
\newblock {FastFlow: Accelerating Deep Learning Model Training with Smart
  Offloading of Input Data Pipeline}.
\newblock {\em Proceedings of the VLDB Endowment}, 2023.

\bibitem{realtime:dac:2017}
Ying Wang, Huawei Li, and Xiaowei Li.
\newblock {Real-time Meets Approximate Computing: An Elastic CNN Inference
  Accelerator with Adaptive Trade-off Between QoS and QoR}.
\newblock In {\em Design Automation Conference (DAC)}, 2017.

\bibitem{nonesparse:dac:2019}
Ying Wang, Shengwen Liang, Huawei Li, and Xiaowei Li.
\newblock {A None-Sparse Inference Accelerator that Distills and Reuses the
  Computation Redundancy in CNNs}.
\newblock In {\em Design Automation Conference (DAC)}, 2019.

\bibitem{wang2022fpganic}
Zeke Wang, Hongjing Huang, Jie Zhang, Fei Wu, and Gustavo Alonso.
\newblock {FpgaNIC: An FPGA-based Versatile 100Gb SmartNIC for GPUs}.
\newblock In {\em USENIX Annual Technical Conference (USENIX ATC)}, 2022.

\bibitem{xilinx_opencl}
{Xilinx}.
\newblock {Xilinx OpenCL Extension}.
\newblock \url{https://xilinx.github.io/XRT/master/html/opencl_extension.html},
  2022.

\bibitem{hls_xilinx}
{Xilinx}.
\newblock {Vitis High-level Synthesis User Guide}.
\newblock \url{https://docs.xilinx.com/r/en-US/ug1399-vitis-hls}, 2023.

\bibitem{FPGAcost}
Xilinx.
\newblock
  \url{https://www.xilinx.com/products/boards-and-kits/alveo/u55c.html}, 2024.

\bibitem{LAX}
Tsung~Tai Yeh, Matthew~D. Sinclair, Bradford~M. Beckmann, and Timothy~G.
  Rogers.
\newblock {Deadline-Aware Offloading for High-Throughput Accelerators}.
\newblock In {\em Proceedings of the International Symposium on
  High-Performance Computer Architecture (HPCA)}, 2021.

\bibitem{yu2022orca}
Gyeong-In Yu, Joo~Seong Jeong, Geon-Woo Kim, Soojeong Kim, and Byung-Gon Chun.
\newblock {Orca: A Distributed Serving System for Transformer-Based Generative
  Models}.
\newblock In {\em USENIX Symposium on Operating Systems Design and
  Implementation (OSDI)}, 2022.

\bibitem{mark}
Chengliang Zhang, Minchen Yu, Wei Wang, and Feng Yan.
\newblock {MArk: Exploiting Cloud Services for Cost-Effective,SLO-Aware Machine
  Learning Inference Serving}.
\newblock In {\em Proceedings of USENIX Annual Technical Conference (ATC)},
  2019.

\bibitem{zhao2022dsi}
Mark Zhao, Niket Agarwal, Aarti Basant, Bu\u{g}ra Gedik, Satadru Pan, Mustafa
  Ozdal, Rakesh Komuravelli, Jerry Pan, Tianshu Bao, Haowei Lu, Sundaram
  Narayanan, Jack Langman, Kevin Wilfong, Harsha Rastogi, Carole-Jean Wu,
  Christos Kozyrakis, and Parik Pol.
\newblock {Understanding Data Storage and Ingestion for Large-Scale Deep
  Recommendation Model Training: Industrial Product}.
\newblock In {\em Proceedings of the International Symposium on Computer
  Architecture (ISCA)}, 2022.

\end{thebibliography}

\end{document}